\documentclass[aps,twocolumn,prd,preprintnumbers, 10pt]{revtex4-2}
\usepackage{xpatch}
\makeatletter
\patchcmd{\@ssect@ltx}
    {\addcontentsline{toc}{#1}{\protect\numberline{}#8}}
    {}
    {}
    {}
\makeatother
\pdfoutput=1
\usepackage{color}
\usepackage{enumitem}
\usepackage[dvipsnames]{xcolor}
\usepackage{hyperref}
\definecolor{amaranth}{rgb}{0.9, 0.17, 0.31}
\definecolor{forestgreen(web)}{rgb}{0.13, 0.55, 0.13}
\definecolor{blue(munsell)}{rgb}{0.0, 0.5, 0.69}
\definecolor{bblue}{rgb}{0.0, 0.58, 0.71}
\hypersetup{
pdfstartview={FitH}, 
pdftitle={}, 
colorlinks=true, 
linkcolor=blue(munsell), 
citecolor=blue(munsell), 
filecolor=magenta, 
urlcolor=blue(munsell)
}
\usepackage{pgfplots}
\usepackage{amssymb}
\usepackage{amsfonts}
\usepackage{graphicx}
\usepackage{epstopdf}
\usepackage{dcolumn}
\usepackage{amsmath}
\usepackage{latexsym,bm}
\usepackage{amsthm}
\usepackage{slashed}
\usepackage{float}
\usepackage{color}
\usepackage{url}
\usepackage{longtable}
\usepackage{rotating}
\usepackage[normalem]{ulem}
\usepackage{faktor}
\usepackage{tikz-cd}
\usepackage{tensor}
\usepackage{array}
\usepackage{tabularx}
\usepackage{makecell}
\usepackage{changepage}

\usepackage{tikz}
\usepackage{diagbox}
\usetikzlibrary{positioning}
\usetikzlibrary{calc}
\usetikzlibrary{decorations.pathreplacing,calligraphy}

\usepackage{xstring}
\usetikzlibrary{decorations.pathmorphing} 
\usetikzlibrary{decorations.markings} 
\usetikzlibrary{arrows} 
\usetikzlibrary{shapes} 
\usetikzlibrary{matrix} 
\usetikzlibrary{positioning} 
\usepackage[english]{babel} 
\usepackage[autostyle]{csquotes}
\usepackage{pifont}
\usetikzlibrary{shapes.multipart}

\tikzset{->-/.style={decoration={
  markings,
  mark=at position .5 with {\arrow{>}}},postaction={decorate}}}

\newcommand{\bea}{\begin{eqnarray}}
\newcommand{\eea}{\end{eqnarray}}
\newcommand{\be}{\begin{equation}}
\newcommand{\ee}{\end{equation}}
\newcommand{\ba}{\begin{aligned}}
\newcommand{\ea}{\end{aligned}}
\newcommand{\bit}{\begin{itemize}}
\newcommand{\eit}{\end{itemize}}
\newcommand{\ben}{\begin{enumerate}}
\newcommand{\een}{\end{enumerate}}

\newcommand{\id}{\text{id}}
\newcommand{\SP}{\text{SP}}

\newcommand{\TSP}{\text{3SP}}

\newcommand{\Bsym}{\mathfrak{B}^{\text{sym}}}
\newcommand{\Bphys}{\mathfrak{B}^{\text{phys}}}

\newcommand{\ot}{\otimes}
\newcommand{\half}{\frac{1}{2}}

\newcommand{\Z}{{\mathbb Z}}

\newcommand{\cA}{\mathcal{A}}

\newcommand{\cC}{\mathcal{C}}

\newcommand{\cE}{\mathcal{E}}

\newcommand{\cH}{\mathcal{H}}
\newcommand{\cI}{\mathcal{I}}

\newcommand{\cK}{\mathcal{K}}
\newcommand{\cL}{\mathcal{L}}
\newcommand{\cM}{\mathcal{M}}

\newcommand{\cO}{\mathcal{O}}

\newcommand{\cQ}{\mathcal{Q}}

\newcommand{\cS}{\mathcal{S}}

\newcommand{\cZ}{\mathcal{Z}}

\newcommand{\fT}{\mathfrak{T}}
\newcommand{\fB}{\mathfrak{B}}

\newcommand{\Hom}{\text{Hom}}

\newcommand{\Mod}{\mathsf{Mod}}

\renewcommand{\dim}{\text{dim}}

\newcommand{\sym}{\text{sym}}
\newcommand{\phys}{\text{phys}}

\makeatother

\begin{document}

\title{A Gapless Phase with Haagerup Symmetry}

\author{Lea E.\ Bottini}
\author{Sakura Sch\"afer-Nameki}

\affiliation{Mathematical Institute, University
of Oxford, Woodstock Road, Oxford, OX2 6GG, United Kingdom}


\begin{abstract} 
\noindent 
We construct a (1+1)d gapless theory which has Haagerup $\mathcal{H}_3$ symmetry. The construction relies on the recent exploration of the categorical Landau paradigm applied to fusion category symmetries. First, using the Symmetry Topological Field Theory, we construct all gapped phases with Haagerup symmetry. Extending this construction to gapless phases, we study the second order phase transition between gapped phases, and determine analytically an Haagerup-symmetric conformal field theory. This is given in terms of two copies of the three-state Potts model, on which we realise the full $\mathcal{H}_3$ symmetry action and determine the relevant deformations to the $\mathcal{H}_3$-symmetric gapped phases. This continuum analysis is corroborated by a lattice model construction of the gapped and gapless phases, using the anyon chain.

\end{abstract}


\maketitle


\noindent{\bf Introduction.}
Symmetries are central to solving many problems in quantum field theory and quantum lattice models. With the advent of generalized \cite{Gaiotto:2014kfa} and non-invertible symmetries, or higher-categorical symmetries \cite{Kaidi:2021xfk, Choi:2021kmx, Bhardwaj:2022yxj} collectively (for reviews see \cite{Schafer-Nameki:2023jdn, Shao:2023gho}), the question of the utility of these new symmetries has been an engine for many exciting developments. However, going back to the 1980s, the importance of fusion category symmetries was already observed in the seminal work  by Moore and Seiberg \cite{Moore:1988qv}. This provides a 
precise mathematical framework to study rational conformal field theories (RCFTs), which naturally come equipped with such symmetries. 
More precisely, the representation category of the vertex operator algebra (VOA) of an RCFT is a modular tensor category (MTC). 

An interesting -- and at present open -- question is whether given any MTC one can construct the associated RCFT.
Despite the existence of endless examples where the RCFTs can be constructed explicitly, some cases remains elusive, the most prominent being the Haagerup fusion category $\cH_3$ \cite{haagerup1994principal, grossman2012quantum}, with the associated MTC being the corresponding Drinfeld center  $\cZ (\cH_3)$ \cite{Izumi:2001mi, Evans:2010yr}. 

A related question is that of finding a 2d conformal field theory (CFT) which has topological lines that form the fusion category $\cH_3$. 
Numerous attempts have been made to construct a gapless theory with Haagerup symmetry $\cH_3$: numerical evidence for a Haagerup-symmetric CFT was obtained using anyon chain models (lattice models with built in fusion category symmetries) with Haagerup symmetry in 
\cite{Huang:2021nvb, Vanhove:2021zop} with central charge $c\sim 2$; subsequently, studies of finite-size effects were performed in \cite{Liu:2022qwn}. 
Most recently,  \cite{Corcoran:2024eeh} constructed an integrable spin chain with broken $\cH_3$ symmetry, as well as a critical spin chain, where $c\sim 3/2$. In summary, the definitive construction -- especially analytically -- of a gapless theory (or CFT) with $\cH_3$ topological lines remains an open question. 

We will not aim to solve this problem here either, at least not under the assumption that the 2d CFT has a single universe/vacuum. However, we will address a very much related question, which is the construction of {\it a} CFT that is $\cH_3$-symmetric. 
We will systematically determine gapped and gapless phases with $\cH_3$ fusion category symmetry and explore the full phase diagram. In particular, we find a two universes/vacua gapless phase with Haagerup symmetry, which admits relevant deformations to Haagerup-symmetric gapped phases. 
The philosophy that we will follow is that of the categorical Landau paradigm \cite{Bhardwaj:2023fca}, which has a systematic implementation for any finite symmetry, in particular those described by fusion categories, via the so-called Symmetry Topological Field Theory (SymTFT) \cite{Ji:2019jhk, Gaiotto:2020iye, Apruzzi:2021nmk, Freed:2022qnc}. For the fusion category symmetry $\cH_3$, the SymTFT has topological defects given by the corresponding Drinfeld center $\cZ(\cH_3)$. Gapped phases are classified in terms of Lagrangian algebras of the center, whereas gapless phases can be studied using the so-called club-sandwich construction in the SymTFT \cite{Bhardwaj:2023bbf, Bhardwaj:2024qrf} (this was discussed for group-like symmetries in \cite{Chatterjee:2022tyg}; see also \cite{Chen:2022wvy} for a holographic approach to phases). This is a generalisation of the standard SymTFT sandwich involving an interface between the topological order $\cZ(\cH_3)$ and a reduced topological order (in the present case $\cZ(\Z_3)$).

\noindent{\bf Summary of results.} We first find all the gapped phases with Haagerup symmetry using the SymTFT approach: there are three gapped phases -- determined previously in \cite{Huang:2021ytb} -- with 2, 4, and 6 vacua, respectively. We then  construct a Haagerup-symmetric {\it gapless} phase that models the second-order phase transition between the 2 and 6 vacua gapped phases. This takes the form of two copies of the three-state Potts model (3SP) with the symmetry $\cH_3$ realized as follows (shown in terms of the arrows, whose precise meaning is detailed in (\ref{eq:symm_gen_Potts})): 
\be\label{bingo}
\begin{tikzpicture}
\node at (1.5,-3) {$\text{\bf 3SP}_0\oplus\text{\bf 3SP}_1$};
\node[right] at (3.2,-3) {$\Z_3, \rho$};
\draw [-stealth](2.7,-3.2) .. controls (3.2,-3.5) and (3.2,-2.6) .. (2.7,-2.9);
\draw [-stealth,rotate=180](-0.3,2.8) .. controls (0.2,2.5) and (0.2,3.4) .. (-0.3,3.1);
\node at (-0.4,-3) {$\Z_3$};
\draw [-stealth](0.8,-3.3) .. controls (1,-3.8) and (2,-3.8) .. (2.2,-3.3);
\node at (1.5,-4) {$\rho$};
\draw [-stealth,rotate=180](-2.2,2.7) .. controls (-2,2.2) and (-1,2.2) .. (-0.8,2.7);
\node at (1.5,-2) {$\rho$};
\end{tikzpicture}
\ee
Here $\Z_3$ refers to the action of the invertible $\Z_3$ sub-symmetry of $\cH_3$, which acts within each of the 3SP models, and $\rho$ is the non-invertible symmetry generator in the Haagerup fusion category, which maps between them. Although not a single universe CFT, this is a fully Haagerup-symmetric 2d CFT.  In particular, the central charge for the model that we find is $c= 4/5$.

\begin{figure}
\centering
\begin{tikzpicture}
\scalebox{0.8}{    \draw[fill= lightgray, opacity=0.2] (0,0) -- (3,0) -- (3,3) -- (0,3) -- (0,0 )  ;
        \draw[fill= lightgray, opacity=0.2] (0,0) -- (-3,0) -- (-3,-3) -- (0,-3) -- (0,0 )  ;
            \draw[fill= lightgray, opacity=0.2] (0,0) -- (3,0) -- (3,3) -- (0,3) -- (0,0 )  ;         
    \draw[fill= lightgray, opacity=0.2] (3,3) -- (3,-3) -- (-3,-3) -- (-3,3) -- (3,3 )  ;
    \draw[white, fill= white] (-3,0.05) -- (3,0.05)-- (3,-0.05) -- (-3, -0.05) -- (-3, 0.05)   ; 
        \draw[white, fill= white] (0.05, -3) -- (0.05, 3)-- (-0.05, 3) -- ( -0.05, -3 ) -- ( 0.05, -3)   ; 
    \draw[->, thick] (-3,0) -- (3,0)     node[right] {$\epsilon_0$}; 
    \draw[thick] (0,-3) -- (0,0);
    \draw[->, thick] (0,0.5) -- (0,3) node[above] {$\epsilon_1$}; 
}
    \node[] at (1.5,1.5) {$\Phi_6^{++}$};
        \node[] at (-1.5,-1.5) {$\Phi_2^{--}$};
            \node[] at (1.5,-1.5) {$\Phi_4^{+-}$};
            \node[] at  (-1.5,1.5) {$\Phi_4^{-+}$};
       \draw[darkgray, dashed] (0.5,0.5) -- (1,1);
       \draw[darkgray, dashed] (0,0) -- (-1,-1);
       \draw[darkgray, dashed] (0,0) -- (1,-1);
       \draw[darkgray, dashed] (-0.5,0.5) -- (-1,1);
    \draw [darkgray,fill=MidnightBlue] (0,0) ellipse (0.09 and 0.09);
    \node[above, MidnightBlue] at (0,0) {\textbf{$\TSP_0 \oplus \TSP_1$}};
\end{tikzpicture}
\caption{Phase diagram of the Haagerup-symmetric theories. We have a critical model at the origin ($\textcolor{MidnightBlue}{\bullet}$),
given by two copies of the three-state Potts model $\TSP_0 \oplus \TSP_1$. This has relevant deformations to the gapped phases $\Phi^{\pm, \pm}_n$, where $(\pm, \pm)$ indicates the sign of the deformation $(\epsilon_0, \epsilon_1)$ and $n$ is the number of vacua in the gapped phase. 
The relevant deformation for each $3 \SP_i$ model gives  $3$ vacua for $\epsilon_i >0$ ($\Z_3$ SSB) and $1$ vacuum for $\epsilon_i<0$ ($\Z_3$ trivial phase).
There is a second order Haagerup-symmetric phase transition between the $\Phi_6^{++}$ and $\Phi_2^{--}$ phases. From  the critical point, there are furthermore deformations to  4 vacua gapped  phases. \label{fig:PhaseDiag} }
\end{figure}

This CFT (\ref{bingo}) admits relevant deformations by $\pm\epsilon_{i}$, $i=0, 1$. Each of the $3\SP_i$ models either flows to a $\Z_3$ spontaneous symmetry breaking (SSB) or trivial phase, depending on the sign of $\epsilon_i$. For the two universe $\cH_3$-symmetric CFT (\ref{bingo}), deformations where $\epsilon_0$ and $\epsilon_1$ have the same sign correspond to relevant deformations to the gapped $\cH_3$-symmetric phases with 2 and 6 vacua. We also can consider deformations with opposite signs $\epsilon_0 =-\epsilon_1$, which flow to a 4 vacua gapped phase. 
The proposed phase diagram is shown in figure \ref{fig:PhaseDiag}. 

We also note that the CFT (\ref{bingo}) can be thought of as the stacking of the 3SP CFT with the Haagerup symmetric TQFT with 2 vacua \footnote{In our construction, the 3SP model could be replaced by any other $\Z_3$-symmetric CFT at the transition between a $\Z_3$ SSB phase and the trivial phase.}.
Of course any CFT times an $\cH_3$-symmetric TQFT would trivially give a gapless phase with $\cH_3$ symmetry. However, the key distinction is that the theory (\ref{bingo}) correctly models the phase transitions between the gapped phases with Haagerup symmetry. 

We corroborate the continuum  analysis by also constructing  the anyon chains that realise the gapped phases, as well as the gapless phase (\ref{bingo}), following the general approach in \cite{Bhardwaj:2024kvy}. This approach is tailored to construct the lattice models directly from the SymTFT and club sandwich data. Similar lattice model constructions for fusion category symmetries have appeared in \cite{Feiguin:2006ydp,2009arXiv0902.3275T, Aasen:2016dop, Buican:2017rxc, Aasen:2020jwb, Lootens:2021tet, Lootens:2022avn, Inamura:2021szw}. 

The general considerations of condensable algebras in the SymTFT furthermore implies that the CFT (\ref{bingo}) should arise as a relevant deformation of the single-universe Haagerup CFT. Thus, we certainly have not solved this specific long-standing problem, but have provided one (and to our knowledge the first concrete proposal for an) RCFT that has Haagerup symmetry, where the symmetry action can be realized concretely and that admits relevant deformations to the gapped phases with Haagerup symmetry. This in itself seems a noteworthy observation, and certainly another constraint on the yet to be uncovered single-universe Haagerup CFT. 
The analysis that we presented here can be applied to any fusion category symmetry and it would be interesting to apply to other setups where CFTs realising that categorical symmetry are unknown \cite{grossman2016brauer,hong2008exotic}. 

\smallskip

\noindent{\bf Haagerup $\cH_3$ Symmetry.}
We consider the Haagerup fusion category $\cH_3$, which has six simple objects
\begin{equation}\label{Baubles}
    \{ 1, \alpha, \alpha^2, \rho, \alpha \rho, \alpha^2 \rho \} \,.
\end{equation}
The non-trivial fusion rules are 
\begin{equation}
    \alpha^3=1 \,, \quad \alpha \rho = \rho \alpha^2 \,, \quad \rho^2 = 1 \oplus \rho \oplus \alpha \rho \oplus \alpha^2 \rho \,.
\end{equation}
In particular, $\{1, \alpha, \alpha^2 \}$ form an invertible $\Z_3$ subcategory, while $\rho$ is the generator of a non-invertible symmetry, with quantum dimension $d = \frac{1}{2} \left(3+\sqrt{13}\right)$. Notice also that due to the non-commutativity of the fusion rules, this category does not admit braiding. 

\smallskip

\noindent{\bf Gapped Phases.}
To study the gapped phases with Haagerup symmetry, we follow  the SymTFT approach outlined in \cite{Bhardwaj:2023ayw, Bhardwaj:2023fca, Bhardwaj:2023idu}. 
The basic idea is that gapped phases are characterized by topological/gapped boundary conditions (BCs) of the 2+1d SymTFT compactified on an interval with two such BCs:
\begin{equation}
\begin{tikzpicture}
\begin{scope}[shift={(0,0)}]
\draw [fill= black, opacity =0.1] (0,-1) -- (0,1)--(2,1) -- (2, -1) -- (0,-1) ; 
\draw [thick] (0,0) -- (2,0);
\draw [thick] (2,-1) -- (2,1);
\draw [thick] (0,-1) -- (0,1);
\node[above ] at (1,0) {$\ell$} ; 
\node at (1, -0.5) {\text{SymTFT}};
\node[above] at (0,1) {$\Bsym$}; 
\node[above] at (2,1) {$\Bphys$}; 
\draw [black,fill=black] (2,0) ellipse (0.05 and 0.05);
\draw [black,fill=black] (0,0) ellipse (0.05 and 0.05);
\end{scope}
\begin{scope}[shift={(4,0)}]
\node at (-1,0) {$=$};
\draw [thick] (0,-1) -- (0,1);
\node[above] at (0,1) {$\text{Gapped Phase}$}; 
\node[right] at (0,0) {$\text{OP}$};
\draw [black,fill=black] (0,0) ellipse (0.05 and 0.05);
\end{scope}
\end{tikzpicture}
\end{equation}
$\Bsym$ specifies the symmetry $\cH_3$ and remains fixed. $\Bphys$ can be any gapped BC and each such choice corresponds to a gapped phase after interval compactification. 

A gapped BC is described in terms of a Lagrangian algebra $\cL$ of the 
Drinfeld center $\cZ(\cH_3)$, physically describing the set of anyons $\ell$ in the SymTFT that can fully terminate on the boundary. Anyons that can terminate on $\Bphys$ characterise the order parameters (OPs) of the gapped phase. Vacua are in correspondence with anyons fully ending on both boundaries.

The simple anyons in $\cZ(\cH_3)$ are given by \cite{Izumi:2001mi, hong2008exotic} 
\be
\{\id, \pi_1, \pi_2, \sigma_1, \sigma_2, \sigma_3, \mu_1, \mu_2, \mu_3, \mu_4, \mu_5, \mu_6 \}\,,
\ee
and we summarize their properties in supplementary materials (SM). 
We determine three irreducible gapped boundary conditions
\be\label{eq:lag}
\ba
\cL_1 &= 1 \oplus \pi_1 \oplus 2 \sigma_1 \\
\cL_2 &=1 \oplus \pi_1 \oplus \pi_2 \oplus \sigma_1\\
\cL_3 &= 1  \oplus \pi_1 \oplus 2 \pi_ 2  \,.
\ea
\ee
The choice $\cL_\sym = \cL_3$ gives the Haagerup $\cH_3$ fusion category symmetry, which we fix for all gapped phases. 
Choosing $\cL_{\text{phys}} = \cL_i$, $i=1,2,3$, gives rise to all the possible gapped phases with $\cH_3$ symmetry. We now discuss the resulting three gapped phases in turn:

\noindent
{\bf $\Z_3$-unbroken Phase $\Phi_{2}^{--}$:}
This corresponds to $\cL_\phys = \cL_1$. The resulting TQFT has two vacua $v_0$ and $v_1$. The action of the $\Z_3$ symmetry generators is trivial on both of them
\begin{equation}
    \alpha: \quad v_0 \rightarrow v_0 \quad,\quad v_1 \rightarrow v_1 \,,
\end{equation}
while the action of the non-invertible generator $\rho$ is \footnote{Notice in particular the presence of non-trivial Euler counter-terms encoded in the factors of $d$ that typically characterise a non-invertible symmetry action in the case where the vacua are physically distinguishable \cite{Bhardwaj:2023idu}.}
\begin{equation}
    \rho: v_0 \rightarrow d^{-1}\, v_1 \quad,\quad v_1 \rightarrow d \,v_0    + 3 v_1 \,,
\end{equation}
This gapped phase can be characterized in terms of the vev of a local operator $\cO_1$ sitting in a multiplet of the $\cH_3$ action labelled by the bulk anyon $\pi_1$
and by the condensation of two purely string order parameters which sit in a multiplet labelled by $\sigma_1$. 

\noindent{ \bf $\Z_3$ SSB $+$ $\Z_3$ Trivial Phase: $\Phi_4^{\pm \mp
}$:}
This corresponds to $\cL_\phys = \cL_2$. The resulting TQFT has four vacua $v_0$, $v_1$, $v_2$ and $v_3$.  The invertible $\Z_3$ symmetry generator $\alpha$  permutes the first three vacua and leaves $v_3$ fixed, so that this phase decomposes as a the sum of a $\Z_3$ SSB phase and a $\Z_3$ trivial phase.
Here $\rho$ acts as 
\begin{equation}
\rho: \quad 
\ba
&v_0 \rightarrow v_0 + a^{-1} v_3\,, \\
    &v_1 \rightarrow v_2 + a^{-1} v_3 \,,
\ea
\quad   
\ba    
    &v_2 \rightarrow v_1 + a^{-1} v_3\,, \\
    &v_3 \rightarrow a (v_0+v_1+v_2) + 2 v_3 \,,
\ea   
\end{equation}
with $a = \frac{1+\sqrt{13}}{2}$. 
This phase can be characterized in terms of the condensation of a local operator $\cO_1$ in the $\pi_1$ multiplet and local operators $\cO_2^1$ and $\cO_2^1$ in the $\pi_2$ multiplet. 
This phase also has a string order parameter labelled by the anyon $\sigma_1$. 

\noindent{\bf $\cH_3$-SSB Phase $\Phi_6^{++}$:} This corresponds to $\cL_\phys = \cL_3$. The resulting TQFT has six vacua $v_i$, $i=1,
\dots,6$. The $\Z_3$ invertible symmetry permutes the first set of vacua $v_0$, $v_1$ and $v_2$ among themselves, as well as the second set of vacua $v_3$, $v_4$ and $v_5$. 
The action of $\rho$ is
\begin{equation}
\begin{aligned}
    \rho: \qquad &v_0 \rightarrow v_0 + v_1 + v_2 + d \,v_3 \\ 
    &v_1 \rightarrow v_0 + v_1 + v_2 + d \,v_5 \\ 
    &v_2 \rightarrow v_0 + v_1 + v_2 + d \,v_4 \\ 
    &v_3 \rightarrow d^{-1} v_0  \,,\ v_4 \rightarrow d^{-1} v_2  \,,\ 
    v_5 \rightarrow d^{-1} v_1  \,,
\end{aligned}    
\end{equation}
The $\cH_3$ symmetry is fully spontaneously broken.
This phase can be characterised in terms of the condensation of the local operators $\cO_1$ in the $\pi_1$ multiplet, $\cO_2^{1,1}$ and $\cO_2^{2,1}$ in the multiplet labelled by  $\pi_2$, and another doublet $\cO_2^{1,2}$ and $\cO_2^{2,2}$ also in the $\pi_2$ multiplet. There are no multiplets involving purely string order parameters condensing in this phase.

\begin{figure}
$$
\begin{tikzpicture}[vertex/.style={draw}]]
\begin{scope}[shift={(0,0)}]
\node[vertex] (1)  at (0,0) {\footnotesize $\cA_0=1$};
\node[vertex] (2) at (0, -1.5) {\footnotesize
\begin{tabular}{c}
$\text{\bf 3SP}_0 \oplus \text{\bf 3SP}_1$\\
{gapless}\\
 $\cA_1  = 1 \oplus \pi_1 $ \\ 
\end{tabular}
} ;
\node[vertex] (3) at (-2.7, -3.5) {\footnotesize 
\begin{tabular}{c}
$\Phi_{2}^{--}$ \\ 
$\Z_3$ trivial\\ 
$\cL_1= 1 \oplus \pi_1 \oplus 2 \sigma_1$  
\end{tabular}
} ;
\node[vertex] (5) at (3, -3.5) {\footnotesize 
\begin{tabular}{c}
$\Phi_{4}^{\pm \mp}$ \\ 
$\Z_3$ SSB $\oplus$ $\Z_3$ trivial\\ 
$\cL_2= 1 \oplus \pi_1 \oplus \pi_2 \oplus \sigma_1 $ 
\end{tabular}
} ;
\node[vertex] (4) at (0, -3.5) {\footnotesize 
\begin{tabular}{c}
$\Phi_6^{++}$\\
$\cH_3$ SSB \\ 
$\cL_3  = 1  \oplus \pi_1  \oplus 2 \pi_ 2$ 
\end{tabular}
};
\draw[-stealth] (1) edge [thick] node[label=left:] {} (2);
\draw[-stealth] (2) edge [thick] node[label=left:] {} (3);
\draw[-stealth] (2) edge [thick] node[label=left:] {} (4);
\draw[-stealth] (1) edge [thick] node[label=left:] {} (5);
\end{scope}
\end{tikzpicture}
$$
\caption{Hasse diagram of gapped and gapless phases for the Haagerup symmetry. The bottom row represents the gapped phases, $\Phi_n$ with  $n$ vacua, and the associated gapped BC (Lagrangian algebra). $\cL_3$ is also the symmetry BC.
The gapless phase we will construct is determined by $\cA_1$, which characterises a phase transition between the gapped phases $\cL_1$ and $\cL_3$. 
The algebra $\cA_0$ represents the universal gapless phases of the Haagerup symmetry. Note that the intersection $\cL_{1} \cap \cL_{2}$ and $\cL_{2} \cap  \cL_{3}$ are not condensable algebras. \label{fig:Hasse}}
\end{figure}

\smallskip

\noindent
{\bf Gapless Phases.}
We now consider the possible gapless phases with Haagerup symmetry. As outlined in \cite{Bhardwaj:2023bbf,Bhardwaj:2024qrf} (see also \cite{Huang:2023pyk,Chatterjee:2022tyg,Wen:2023otf} for group-like cases), every condensable algebra in $\cZ(\cS)$ defines a $\cS$-symmetric phase, with Lagrangian algebras corresponding to gapped phases and non-Lagrangian condensable algebras to gapless phases. Moreover, condensable algebras can be arranged into a Hasse diagram where the partial order is given by inclusion, i.e.\ $\cA_1 < \cA_2$ if $\cA_1$ is a subalgebra of $\cA_2$ \cite{Bhardwaj:2024qrf}. This is shown in figure \ref{fig:Hasse} for the Haagerup symmetry.

We find that there is only one non-trivial non-Lagrangian algebra in $\cZ(\cH_3)$, namely we have 
\be
\cA_0 = 1 \,,\qquad 
\cA_1  = 1 \oplus \pi_1  \,. 
\ee
First of all, notice that $\cA_1 = \cL_3 \cap \cL_1$, so that we expect the gapless phase corresponding to $\cA_1$ to be a standard critical point describing the phase transition between the 6 vacua gapped phase and the 2 vacua gapped phase, specified by the Lagrangian algebras $\cL_3$ and $\cL_1$ respectively. Second, notice that $\cA_1$ is not a subalgebra of $\cL_2$, which has only the trivial $\cA_0$.\footnote{We thank Corey Jones and the authors of \cite{Hung:2025gcp} for pointing this out.} \footnote{Moreover, notice that the other intersections between the Lagrangian algebras, namely $\cL_1 \cap \cL_2 = 1 \oplus \pi_1 \oplus \sigma_1$ and $\cL_3 \cap \cL_2 = 1 \oplus \pi_1 \oplus \pi_2$, are not condensable. Thus from the SymTFT point of view we can only infer the existence of a second order Haagerup symmetric transition between the 2 and 6 vacua gapped phases, but not between the 6 and the 4 vacua gapped phases (and similarly between 4 and 2).}

We now recall how to characterize $\cS$-symmetric phase transitions using the club sandwich introduced in \cite{Bhardwaj:2023bbf}. A condensable algebra defines an interface between $\cZ(\cS)$ and a reduced topological order $\cZ(\cS')$:
\be\label{eq:club_sandwich}
\begin{tikzpicture}
\begin{scope}[shift={(0,0)}]
\draw [fill= black, opacity =0.1] (0,0.6) -- (0,2) -- (2,2) -- (2,0.6) -- (0,0.6)  ; 
\draw [fill= black, opacity =0.1] (4,0.6) -- (4,2) -- (2,2) -- (2,0.6) -- (4,0.6) ; 
\draw [very thick] (0,0.6) -- (0,2) ;
\draw [very thick] (2,0.6) -- (2,2) ;
\draw [very thick] (4,0.6) -- (4,2) ;
\node at (1,1.3) {$\cZ(\cS)$} ;
\node at (3,1.3) {$\cZ(\cS')$} ;
\node[above] at (0,2) {$\Bsym_\cS$}; 
\node[above] at (2,2) {$\cI$}; 
\node[above] at (4,2) {$\Bphys$};
\end{scope}
\node at (4.7,1.3) {=};
\begin{scope}[shift={(3.4,0)}]
\draw [very thick] (2,0.6) -- (2,2) ;
\node[above] at (2,2) {$\fT$}; 
\end{scope}
\draw [thick,-stealth](5.7,2.2) .. controls (6.2,1.8) and (6.2,2.9) .. (5.7,2.5);
\node at (6.4,2.4) {$\cS$};
\end{tikzpicture}
\ee
The input physical boundary $\Bphys$ is a physical boundary for $\cZ(\cS')$, and thus $\cS'$ symmetric to start with. 
The key point of the construction is that the collapse of the full club sandwich yields a theory that is $\cS$-symmetric, thus providing a map from $\cS'$-symmetric theories to $\cS$-symmetric theories. These maps are also known as Kennedy-Tasaki (KT) transformations \cite{kennedy1992hidden, Kennedy:1992ifl}.
This is particularly useful if we want to determine the phase transition between two $\cS$-symmetric gapped phases  $\fT_1$ and  $\fT_2$. 
Let $\mathfrak{T}'_{12}$ be the second order phase transition between two $\cS'$-symmetric gapped phases $\fT'_1$ and $\fT'_2$. This means that there exists a relevant deformation $\epsilon'$ of the CFT such that by deforming the theory with $\pm\epsilon'$ we flow to the two gapped phases:
\be\label{PT1}
\begin{tikzpicture}
\node (v1) at (1,-0.5) {$\mathfrak{T}'_{12}$};
\node (v3) at (3.5,-0.5) {$\fT'_1$};
\node (v2) at (-1.5,-0.5) {$\fT'_2$};
\draw [-stealth] (v1) edge (v2);
\draw [-stealth] (v1) edge (v3);
\node at (-0.2,-0.2) {$-\epsilon'$};
\node at (2.2,-0.2) {$+\epsilon'$};
\end{tikzpicture}
\ee
Inputting $\mathfrak{T}'_{12}$ as the physical boundary $\Bphys$ of the club sandwich, we can use the $\cS' \rightarrow \cS$ map to upgrade $\mathfrak{T}'_{12}$ to a $\cS$-symmetric CFT $\mathfrak{T}_{12}$ that is the second order transition between $\fT_1$ and $\fT_2$:
\be\label{PT1}
\begin{tikzpicture}
\node (v1) at (1,-0.5) {$\mathfrak{T}_{12}$};
\node (v3) at (3.5,-0.5) {$\fT_1$};
\node (v2) at (-1.5,-0.5) {$\fT_2$};
\draw [-stealth] (v1) edge (v2);
\draw [-stealth] (v1) edge (v3);
\node at (-0.2,-0.2) {$-\epsilon$};
\node at (2.2,-0.2) {$+\epsilon$};
\end{tikzpicture}
\ee
This allows us to derive phase transitions for larger symmetries by inputting known phase transitions for smaller, “elementary" symmetries. 

We apply this to the Haagerup $\cH_3$ symmetry, for which there is one non-maximal non-trivial condensable algebra $\cA_{1}= 1 \oplus \pi_1$. The reduced topological order in this case is $\cZ(\Z_3)$. The club sandwich takes the following form
\begin{equation}
\label{CSA1}
\begin{tikzpicture}
\begin{scope}[shift={(0,0)}]
\draw [fill= black, opacity =0.1] (0,-1) -- (0,2)--(4,2) -- (4, -1) -- (0,-1) ; 
\draw [very thick] (0,-1) -- (0,2) ;
\draw [very thick] (2,-1) -- (2,2) ;
\draw [very thick] (4,-1) -- (4,2) ;
\draw[thick] (0,1) -- (4,1);
\draw[thick] (0,0.2) -- (4,0.2);
\draw[thick] (0,-0.5) -- (2,-0.5);
\node at (1,1.65) {$\cZ(\cH_3)$} ;
\node at (3,1.65) {$\cZ(\Z_3)$} ;
\node[above] at (0,2) {$\cL_3$}; 
\node[above] at (2,2) {$\cA_1$}; 
\node[above] at (4,2) {3SP}; 
\node[below] at (1,1) {$\pi_2$}; 
\node[below] at (3,1) {$e^2$}; 
\node[below] at (1,0.2) {$\pi_2$}; 
\node[below] at (3,0.2) {$e$}; 
\node[below] at (1,-0.5) {$\pi_1$}; 
\draw [black,fill=black] (0,1) ellipse (0.05 and 0.05);
\draw [black,fill=black] (0,0.2) ellipse (0.05 and 0.05);
\draw [black,fill=black] (0,-0.5) ellipse (0.05 and 0.05);
\draw [black,fill=black] (2,1) ellipse (0.05 and 0.05);
\draw [black,fill=black] (2,0.2) ellipse (0.05 and 0.05);
\draw [black,fill=black] (2,-0.5) ellipse (0.05 and 0.05);
\node at (4.4,0.5) {$=$};
\end{scope}
\begin{scope}[shift={(5.5,0)}]
\draw [fill= black, opacity =0.1] (0,-1) -- (0,2)--(2,2) -- (2, -1) -- (0,-1) ; 
\draw [very thick] (0,-1) -- (0,2) ;
\draw [very thick] (2,-1) -- (2,2) ;
\node[below] at (1,1) {$e^2$}; 
\node[below] at (1,0.2) {$e$}; 
\node at (1,1.65) {$\cZ(\Z_3)$} ;
\node[above] at (0,2) {$2 \,\cL_e$}; 
\node[above] at (2,2) {3SP}; 
\draw[thick] (0,1) -- (2,1);
\draw[thick] (0,0.2) -- (2,0.2);
\draw [black,fill=black] (0,1) ellipse (0.05 and 0.05);
\draw [black,fill=black] (0,0.2) ellipse (0.05 and 0.05);
\draw [black,fill=black] (0,-0.5) ellipse (0.05 and 0.05);
\end{scope}
\end{tikzpicture}
\end{equation}
Compactifying the left half, the irreducible BC $\cL_3$ for $\cZ(\cH_3)$ becomes the reducible BC $2 \,\cL_e$ for $\cZ(\Z_3)$. We now input a $\Z_3$-symmetric phase transition, e.g.\ the 3-state Potts model (3SP). 
The BC $2 \,\cL_e$ implies that after further compactifying we get two copies of the 3SP, $3\SP_0 \oplus 3\SP_1$, constituting a CFT with two universes which are dynamically decoupled but connected by the action of the $\cH_3$ symmetry. We denote by $\eta_{ii}$, $\eta_{ii}^3 = 1_{ii}$, the generator of the $\Z_3$ symmetry within each $3\SP_i$, and by $1_{ij}$, with $i\neq j$, a universe changing line operator. We derive in the SM that $\cH_3$ acts on  $3\SP_0 \oplus 3\SP_1$ as 
\begin{equation}\label{eq:symm_gen_Potts}
\begin{aligned}
    &\alpha = \eta_{00} \oplus \eta_{11}^2 \\ 
    &\rho = 1_{01} \oplus 1_{10} \oplus 1_{11} \oplus \eta_{11} \oplus \eta^2_{11}\,.
\end{aligned}    
\end{equation}
Each $3\SP_i$ model, $i=0,1$, admits a relevant deformation $\epsilon_i$ making it flow to a three vacua gapped phase for $\epsilon_i >0 $ and to a single vacuum gapped phase for $\epsilon_i <0$. 
Overall, we find the phase diagram in figure \ref{fig:PhaseDiag}. Starting with the theory (\ref{bingo}), all combinations of relevant deformations are allowed and give rise to various gapped phases with $n=2, 4, 6$ vacua, labelled $\Phi_n^{\text{sign}(\epsilon_0), \text{sign}(\epsilon_1)}$. 
The transition between $\Phi_2^{--}$ and $\Phi_6^{++}$ is second order, likewise between $\Phi_4^{+-}$ and $\Phi_4^{-+}$. All the transitions are $\Z_3$ symmetric. Moreover, the $\Phi_2^{--}$ to $\Phi_6^{++}$ transition arises from the club-sandwich and is therefore fully $\cH_3$ symmetric. One can indeed check using \eqref{eq:symm_gen_Potts} and taking into accounts the Euler counter-terms (see SM) that 
$\alpha(\epsilon_0 + \epsilon_1) = \epsilon_0 + \epsilon_1$ and $\rho(\epsilon_0 + \epsilon_1) = d (\epsilon_0 + \epsilon_1)$. This approach does not say anything directly about the 4 vacua gapped phases. 
However, it is very suggestive that the CFT (\ref{bingo}) admits a deformation to a 4  vacua gapped phase by taking $\epsilon_0=-\epsilon_1$, which decomposes as a $\Z_3$ SSB phase and a trivial phase with respect to the symmetry action of $\alpha$.  

\noindent{\bf Lattice model.} We have shown that the theory (\ref{bingo}) admits both the action of $\cH_3$ and relevant deformations to $\cH_3$ symmetric gapped phases, and thus is a CFT with this symmetry. Our analysis was entirely based on the continuum SymTFT description. To complement this, we also applied the general anyon chain construction for gapped and gapless phases of \cite{Bhardwaj:2024kvy} to the $\cH_3$ symmetry in the SM. We construct the Hamiltonians whose ground states realize the three gapped phases, as well as the second-order phase transition. In this case the model decomposes into two critical 3SP spin chains, confirming our continuum results. 

\noindent{\bf Outlook.} We presented the first gapless (1+1)d theory with $\cH_3$ symmetry. Although this is a two-universe CFT, which is in fact a tensor product of the $\cH_3$-symmetric gapped phase with two vacua and the 3SP model, we expect this to have potentially interesting consequences in determining the single universe Haagerup CFT. E.g.\ one could try to apply constraints from modular bootstrap to access this single-universe model. A conjecture motivated by the present work is \footnote{We thank Kantaro Ohmori for discussions on this.}  that the single-universe CFT could be the $c=1$ compact boson times the 3SP model. This would be compatible with the numerical results on $c \sim 2$ in \cite{Huang:2021nvb, Vanhove:2021zop}, but it remains to be seen how the full $\cH_3$ symmetry acts. 
Our approach is very general and applicable to any fusion category symmetry $\cS$, and should shed light on those missing entries in the list of $\cS$-symmetric CFTs.

\noindent{\bf Acknowledgments.}
We thank Andrea Antinucci, Christian Copetti, Luisa Eck, Paul Fendley, Yuhan Gai, Sheng-Jie Huang, Kansei Inamura, Mark Mezei, Kantaro Ohmori, Yuji Tachikawa and Jingxiang Wu for discussions at various stages of this work. 
The work of SSN is supported by the UKRI Frontier Research Grant, underwriting the ERC Advanced Grant "Generalized Symmetries in Quantum Field Theory and Quantum Gravity”.

\phantom{\cite{Huang:2020lox,osborne2019f, haagerup1994principal,asaeda1999exotic, grossman2012quantum, Izumi:2001mi, hong2008exotic, Huang:2021ytb, Lin:2022dhv,Bhardwaj:2023ayw,Bartsch:2023wvv, Ng:2023wsc, Chang:2018iay, Fendley2019, Feiguin:2006ydp,2009arXiv0902.3275T, Aasen:2016dop, Buican:2017rxc, Aasen:2020jwb, Lootens:2021tet, Lootens:2022avn, Inamura:2021szw, Bhardwaj:2024kvy, Bhardwaj:2024kvy, Inamura:2021szw}}

 \bibliographystyle{ytphys}
 \small 
 \baselineskip=.94\baselineskip
 \let\bbb\bibitem\def\bibitem{\itemsep4pt\bbb}
\bibliography{ref.bib}

\providecommand{\href}[2]{#2}\begingroup\raggedright\begin{thebibliography}{10}

\bibitem{Gaiotto:2014kfa}
D.~Gaiotto, A.~Kapustin, N.~Seiberg, and B.~Willett, ``{Generalized Global
  Symmetries},'' \href{http://dx.doi.org/10.1007/JHEP02(2015)172}{{\em JHEP}
  {\bfseries 02} (2015) 172}, \href{http://arxiv.org/abs/1412.5148}{{\ttfamily
  arXiv:1412.5148 [hep-th]}}.

\bibitem{Kaidi:2021xfk}
J.~Kaidi, K.~Ohmori, and Y.~Zheng, ``{Kramers-Wannier-like Duality Defects in
  (3+1)D Gauge Theories},''
  \href{http://dx.doi.org/10.1103/PhysRevLett.128.111601}{{\em Phys. Rev.
  Lett.} {\bfseries 128} no.~11, (2022) 111601},
  \href{http://arxiv.org/abs/2111.01141}{{\ttfamily arXiv:2111.01141
  [hep-th]}}.

\bibitem{Choi:2021kmx}
Y.~Choi, C.~Cordova, P.-S. Hsin, H.~T. Lam, and S.-H. Shao, ``{Noninvertible
  duality defects in 3+1 dimensions},''
  \href{http://dx.doi.org/10.1103/PhysRevD.105.125016}{{\em Phys. Rev. D}
  {\bfseries 105} no.~12, (2022) 125016},
  \href{http://arxiv.org/abs/2111.01139}{{\ttfamily arXiv:2111.01139
  [hep-th]}}.

\bibitem{Bhardwaj:2022yxj}
L.~Bhardwaj, L.~E. Bottini, S.~Schafer-Nameki, and A.~Tiwari, ``{Non-Invertible
  Higher-Categorical Symmetries},''
  \href{http://dx.doi.org/10.21468/SciPostPhys.14.1.007}{{\em SciPost Phys.}
  {\bfseries 14} (2023) 007}, \href{http://arxiv.org/abs/2204.06564}{{\ttfamily
  arXiv:2204.06564 [hep-th]}}.

\bibitem{Schafer-Nameki:2023jdn}
S.~Schafer-Nameki, ``{ICTP Lectures on (Non-)Invertible Generalized
  Symmetries},'' \href{http://arxiv.org/abs/2305.18296}{{\ttfamily
  arXiv:2305.18296 [hep-th]}}.

\bibitem{Shao:2023gho}
S.-H. Shao, ``{What's Done Cannot Be Undone: TASI Lectures on Non-Invertible
  Symmetry},'' \href{http://arxiv.org/abs/2308.00747}{{\ttfamily
  arXiv:2308.00747 [hep-th]}}.

\bibitem{Moore:1988qv}
G.~W. Moore and N.~Seiberg, ``{Classical and Quantum Conformal Field Theory},''
  \href{http://dx.doi.org/10.1007/BF01238857}{{\em Commun. Math. Phys.}
  {\bfseries 123} (1989) 177}.

\bibitem{haagerup1994principal}
U.~Haagerup, ``{Principal graphs of subfactors in the index range $4< [M:N] <
  3+ \sqrt{2}$},'' {\em Subfactors (Proceedings of Taniguchi Symposium,
  Kyuzeso, 1993), pp. 1–38} {\bfseries 1} (1994) .

\bibitem{grossman2012quantum}
P.~Grossman and N.~Snyder, ``{Quantum subgroups of the Haagerup fusion
  categories},'' {\em Communications in Mathematical Physics} {\bfseries 311}
  (2012) 617--643.

\bibitem{Izumi:2001mi}
M.~Izumi, ``{The structure of sectors associated with Longo-Rehren inclusions.
  II: Examples},'' \href{http://dx.doi.org/10.1142/S0129055X01000818}{{\em Rev.
  Math. Phys.} {\bfseries 13} (2001) 603--674}.

\bibitem{Evans:2010yr}
D.~E. Evans and T.~Gannon, ``{The exoticness and realisability of twisted
  Haagerup-Izumi modular data},''
  \href{http://dx.doi.org/10.1007/s00220-011-1329-3}{{\em Commun. Math. Phys.}
  {\bfseries 307} (2011) 463--512},
  \href{http://arxiv.org/abs/1006.1326}{{\ttfamily arXiv:1006.1326 [math.QA]}}.

\bibitem{Huang:2021nvb}
T.-C. Huang, Y.-H. Lin, K.~Ohmori, Y.~Tachikawa, and M.~Tezuka, ``{Numerical
  Evidence for a Haagerup Conformal Field Theory},''
  \href{http://dx.doi.org/10.1103/PhysRevLett.128.231603}{{\em Phys. Rev.
  Lett.} {\bfseries 128} no.~23, (2022) 231603},
  \href{http://arxiv.org/abs/2110.03008}{{\ttfamily arXiv:2110.03008
  [cond-mat.stat-mech]}}.

\bibitem{Vanhove:2021zop}
R.~Vanhove, L.~Lootens, M.~Van~Damme, R.~Wolf, T.~J. Osborne, J.~Haegeman, and
  F.~Verstraete, ``{Critical Lattice Model for a Haagerup Conformal Field
  Theory},'' \href{http://dx.doi.org/10.1103/PhysRevLett.128.231602}{{\em Phys.
  Rev. Lett.} {\bfseries 128} no.~23, (2022) 231602},
  \href{http://arxiv.org/abs/2110.03532}{{\ttfamily arXiv:2110.03532
  [cond-mat.stat-mech]}}.

\bibitem{Liu:2022qwn}
Y.~Liu, Y.~Zou, and S.~Ryu, ``{Operator fusion from wave-function overlap:
  Universal finite-size corrections and application to the Haagerup model},''
  \href{http://dx.doi.org/10.1103/PhysRevB.107.155124}{{\em Phys. Rev. B}
  {\bfseries 107} no.~15, (2023) 155124},
  \href{http://arxiv.org/abs/2203.14992}{{\ttfamily arXiv:2203.14992
  [cond-mat.str-el]}}.

\bibitem{Corcoran:2024eeh}
L.~Corcoran and M.~de~Leeuw, ``{Integrable and critical Haagerup spin
  chains},'' \href{http://arxiv.org/abs/2410.16356}{{\ttfamily arXiv:2410.16356
  [cond-mat.stat-mech]}}.

\bibitem{Bhardwaj:2023fca}
L.~Bhardwaj, L.~E. Bottini, D.~Pajer, and S.~Schafer-Nameki, ``{Categorical
  Landau Paradigm for Gapped Phases},''
  \href{http://arxiv.org/abs/2310.03786}{{\ttfamily arXiv:2310.03786
  [cond-mat.str-el]}}.

\bibitem{Ji:2019jhk}
W.~Ji and X.-G. Wen, ``{Categorical symmetry and noninvertible anomaly in
  symmetry-breaking and topological phase transitions},''
  \href{http://dx.doi.org/10.1103/PhysRevResearch.2.033417}{{\em Phys. Rev.
  Res.} {\bfseries 2} no.~3, (2020) 033417},
  \href{http://arxiv.org/abs/1912.13492}{{\ttfamily arXiv:1912.13492
  [cond-mat.str-el]}}.

\bibitem{Gaiotto:2020iye}
D.~Gaiotto and J.~Kulp, ``{Orbifold groupoids},''
  \href{http://dx.doi.org/10.1007/JHEP02(2021)132}{{\em JHEP} {\bfseries 02}
  (2021) 132}, \href{http://arxiv.org/abs/2008.05960}{{\ttfamily
  arXiv:2008.05960 [hep-th]}}.

\bibitem{Apruzzi:2021nmk}
F.~Apruzzi, F.~Bonetti, I.~n.~G. Etxebarria, S.~S. Hosseini, and
  S.~Schafer-Nameki, ``{Symmetry TFTs from String Theory},''
  \href{http://arxiv.org/abs/2112.02092}{{\ttfamily arXiv:2112.02092
  [hep-th]}}.

\bibitem{Freed:2022qnc}
D.~S. Freed, G.~W. Moore, and C.~Teleman, ``{Topological symmetry in quantum
  field theory},'' \href{http://arxiv.org/abs/2209.07471}{{\ttfamily
  arXiv:2209.07471 [hep-th]}}.

\bibitem{Bhardwaj:2023bbf}
L.~Bhardwaj, L.~E. Bottini, D.~Pajer, and S.~Schafer-Nameki, ``{The Club
  Sandwich: Gapless Phases and Phase Transitions with Non-Invertible
  Symmetries},'' \href{http://arxiv.org/abs/2312.17322}{{\ttfamily
  arXiv:2312.17322 [hep-th]}}.

\bibitem{Bhardwaj:2024qrf}
L.~Bhardwaj, D.~Pajer, S.~Schafer-Nameki, and A.~Warman, ``{Hasse Diagrams for
  Gapless SPT and SSB Phases with Non-Invertible Symmetries},''
  \href{http://arxiv.org/abs/2403.00905}{{\ttfamily arXiv:2403.00905
  [cond-mat.str-el]}}.

\bibitem{Chatterjee:2022tyg}
A.~Chatterjee and X.-G. Wen, ``{Holographic theory for continuous phase
  transitions: Emergence and symmetry protection of gaplessness},''
  \href{http://dx.doi.org/10.1103/PhysRevB.108.075105}{{\em Phys. Rev. B}
  {\bfseries 108} no.~7, (2023) 075105},
  \href{http://arxiv.org/abs/2205.06244}{{\ttfamily arXiv:2205.06244
  [cond-mat.str-el]}}.

\bibitem{Chen:2022wvy}
L.~Chen, H.~Zhang, K.~Ji, C.~Shen, R.~Wang, X.~Zeng, and L.-Y. Hung, ``{CFT$_D$
  from TQFT$_{D+1}$ via Holographic Tensor Network, and Precision
  Discretisation of CFT$_2$},''
  \href{http://arxiv.org/abs/2210.12127}{{\ttfamily arXiv:2210.12127
  [hep-th]}}.

\bibitem{Huang:2021ytb}
T.-C. Huang and Y.-H. Lin, ``{Topological field theory with Haagerup
  symmetry},'' \href{http://dx.doi.org/10.1063/5.0079062}{{\em J. Math. Phys.}
  {\bfseries 63} no.~4, (2022) 042306},
  \href{http://arxiv.org/abs/2102.05664}{{\ttfamily arXiv:2102.05664
  [hep-th]}}.

\bibitem{Note1}
In our construction, the 3SP model could be replaced by any other ${\protect
  \mathbb Z}_3$-symmetric CFT at the transition between a ${\protect \mathbb
  Z}_3$ SSB phase and the trivial phase.

\bibitem{Bhardwaj:2024kvy}
L.~Bhardwaj, L.~E. Bottini, S.~Schafer-Nameki, and A.~Tiwari, ``{Lattice Models
  for Phases and Transitions with Non-Invertible Symmetries},''
  \href{http://arxiv.org/abs/2405.05964}{{\ttfamily arXiv:2405.05964
  [cond-mat.str-el]}}.

\bibitem{Feiguin:2006ydp}
A.~Feiguin, S.~Trebst, A.~W.~W. Ludwig, M.~Troyer, A.~Kitaev, Z.~Wang, and
  M.~H. Freedman, ``{Interacting anyons in topological quantum liquids: The
  golden chain},'' \href{http://dx.doi.org/10.1103/PhysRevLett.98.160409}{{\em
  Phys. Rev. Lett.} {\bfseries 98} (2007) 160409},
  \href{http://arxiv.org/abs/cond-mat/0612341}{{\ttfamily
  arXiv:cond-mat/0612341}}.

\bibitem{2009arXiv0902.3275T}
S.~{Trebst}, M.~{Troyer}, Z.~{Wang}, and A.~W.~W. {Ludwig}, ``{A short
  introduction to Fibonacci anyon models},''
  \href{http://dx.doi.org/10.48550/arXiv.0902.3275}{{\em arXiv e-prints} (Feb.,
  2009) arXiv:0902.3275}, \href{http://arxiv.org/abs/0902.3275}{{\ttfamily
  arXiv:0902.3275 [cond-mat.stat-mech]}}.

\bibitem{Aasen:2016dop}
D.~Aasen, R.~S.~K. Mong, and P.~Fendley, ``{Topological Defects on the Lattice
  I: The Ising model},''
  \href{http://dx.doi.org/10.1088/1751-8113/49/35/354001}{{\em J. Phys. A}
  {\bfseries 49} no.~35, (2016) 354001},
  \href{http://arxiv.org/abs/1601.07185}{{\ttfamily arXiv:1601.07185
  [cond-mat.stat-mech]}}.

\bibitem{Buican:2017rxc}
M.~Buican and A.~Gromov, ``{Anyonic Chains, Topological Defects, and Conformal
  Field Theory},'' \href{http://dx.doi.org/10.1007/s00220-017-2995-6}{{\em
  Commun. Math. Phys.} {\bfseries 356} no.~3, (2017) 1017--1056},
  \href{http://arxiv.org/abs/1701.02800}{{\ttfamily arXiv:1701.02800
  [hep-th]}}.

\bibitem{Aasen:2020jwb}
D.~Aasen, P.~Fendley, and R.~S.~K. Mong, ``{Topological Defects on the Lattice:
  Dualities and Degeneracies},''
  \href{http://arxiv.org/abs/2008.08598}{{\ttfamily arXiv:2008.08598
  [cond-mat.stat-mech]}}.

\bibitem{Lootens:2021tet}
L.~Lootens, C.~Delcamp, G.~Ortiz, and F.~Verstraete, ``{Dualities in
  One-Dimensional Quantum Lattice Models: Symmetric Hamiltonians and Matrix
  Product Operator Intertwiners},''
  \href{http://dx.doi.org/10.1103/PRXQuantum.4.020357}{{\em PRX Quantum}
  {\bfseries 4} no.~2, (2023) 020357},
  \href{http://arxiv.org/abs/2112.09091}{{\ttfamily arXiv:2112.09091
  [quant-ph]}}.

\bibitem{Lootens:2022avn}
L.~Lootens, C.~Delcamp, and F.~Verstraete, ``{Dualities in One-Dimensional
  Quantum Lattice Models: Topological Sectors},''
  \href{http://dx.doi.org/10.1103/PRXQuantum.5.010338}{{\em PRX Quantum}
  {\bfseries 5} no.~1, (2024) 010338},
  \href{http://arxiv.org/abs/2211.03777}{{\ttfamily arXiv:2211.03777
  [quant-ph]}}.

\bibitem{Inamura:2021szw}
K.~Inamura, ``{On lattice models of gapped phases with fusion category
  symmetries},'' \href{http://dx.doi.org/10.1007/JHEP03(2022)036}{{\em JHEP}
  {\bfseries 03} (2022) 036}, \href{http://arxiv.org/abs/2110.12882}{{\ttfamily
  arXiv:2110.12882 [cond-mat.str-el]}}.

\bibitem{grossman2016brauer}
P.~Grossman and N.~Snyder, ``{The Brauer-Picard group of the Asaeda-Haagerup
  fusion categories},'' {\em Transactions of the American Mathematical Society}
  {\bfseries 368} no.~4, (2016) 2289--2331.

\bibitem{hong2008exotic}
S.-M. Hong, E.~Rowell, and Z.~Wang, ``On exotic modular tensor categories,''
  {\em Communications in Contemporary Mathematics} {\bfseries 10} no.~supp01,
  (2008) 1049--1074.

\bibitem{Bhardwaj:2023ayw}
L.~Bhardwaj and S.~Schafer-Nameki, ``{Generalized Charges, Part II:
  Non-Invertible Symmetries and the Symmetry TFT},''
  \href{http://arxiv.org/abs/2305.17159}{{\ttfamily arXiv:2305.17159
  [hep-th]}}.

\bibitem{Bhardwaj:2023idu}
L.~Bhardwaj, L.~E. Bottini, D.~Pajer, and S.~Schafer-Nameki, ``{Gapped Phases
  with Non-Invertible Symmetries: (1+1)d},''
  \href{http://arxiv.org/abs/2310.03784}{{\ttfamily arXiv:2310.03784
  [hep-th]}}.

\bibitem{Note2}
Notice in particular the presence of non-trivial Euler counter-terms encoded in
  the factors of $d$ that typically characterise a non-invertible symmetry
  action in the case where the vacua are physically distinguishable \cite
  {Bhardwaj:2023idu}.

\bibitem{Huang:2023pyk}
S.-J. Huang and M.~Cheng, ``{Topological holography, quantum criticality, and
  boundary states},'' \href{http://arxiv.org/abs/2310.16878}{{\ttfamily
  arXiv:2310.16878 [cond-mat.str-el]}}.

\bibitem{Wen:2023otf}
R.~Wen and A.~C. Potter, ``{Classification of 1+1D gapless symmetry protected
  phases via topological holography},''
  \href{http://arxiv.org/abs/2311.00050}{{\ttfamily arXiv:2311.00050
  [cond-mat.str-el]}}.

\bibitem{Note3}
We thank Corey Jones and the authors of \cite {Hung:2025gcp} for pointing this
  out.

\bibitem{Note4}
Moreover, notice that the other intersections between the Lagrangian algebras,
  namely $\protect \mathcal {L}_1 \cap \protect \mathcal {L}_2 = 1 \oplus \pi
  _1 \oplus \sigma _1$ and $\protect \mathcal {L}_3 \cap \protect \mathcal
  {L}_2 = 1 \oplus \pi _1 \oplus \pi _2$, are not condensable. Thus from the
  SymTFT point of view we can only infer the existence of a second order
  Haagerup symmetric transition between the 2 and 6 vacua gapped phases, but
  not between the 6 and the 4 vacua gapped phases (and similarly between 4 and
  2).

\bibitem{kennedy1992hidden}
T.~Kennedy and H.~Tasaki, ``Hidden symmetry breaking and the haldane phase in
  s= 1 quantum spin chains,'' {\em Communications in mathematical physics}
  {\bfseries 147} (1992) 431--484.

\bibitem{Kennedy:1992ifl}
T.~Kennedy and H.~Tasaki, ``{Hidden Z2\texttimes{}Z2 symmetry breaking in
  Haldane-gap antiferromagnets},''
  \href{http://dx.doi.org/10.1103/PhysRevB.45.304}{{\em Phys. Rev. B}
  {\bfseries 45} no.~1, (1992) 304}.

\bibitem{Note5}
We thank Kantaro Ohmori for discussions on this.

\bibitem{Huang:2020lox}
T.-C. Huang and Y.-H. Lin, ``{The $F$-Symbols for Transparent Haagerup-Izumi
  Categories with $G = \mathbb{Z}_{2n+1}$},''
  \href{http://arxiv.org/abs/2007.00670}{{\ttfamily arXiv:2007.00670
  [math.CT]}}.

\bibitem{osborne2019f}
T.~J. Osborne, D.~E. Stiegemann, and R.~Wolf, ``{The F-symbols for the $H_3$
  fusion category},'' {\em arXiv preprint arXiv:1906.01322} (2019) .

\bibitem{asaeda1999exotic}
M.~Asaeda and U.~Haagerup, ``{Exotic subfactors of finite depth with Jones
  indices $(5+\sqrt{13})/2$ and $(5+\sqrt{17})/2$},'' {\em Communications in
  mathematical physics} {\bfseries 202} no.~1, (1999) 1--63.

\bibitem{Lin:2022dhv}
Y.-H. Lin, M.~Okada, S.~Seifnashri, and Y.~Tachikawa, ``{Asymptotic density of
  states in 2d CFTs with non-invertible symmetries},''
  \href{http://dx.doi.org/10.1007/JHEP03(2023)094}{{\em JHEP} {\bfseries 03}
  (2023) 094}, \href{http://arxiv.org/abs/2208.05495}{{\ttfamily
  arXiv:2208.05495 [hep-th]}}.

\bibitem{Bartsch:2023wvv}
T.~Bartsch, M.~Bullimore, and A.~Grigoletto, ``{Representation theory for
  categorical symmetries},'' \href{http://arxiv.org/abs/2305.17165}{{\ttfamily
  arXiv:2305.17165 [hep-th]}}.

\bibitem{Ng:2023wsc}
S.-H. Ng, E.~C. Rowell, and X.-G. Wen, ``{Classification of modular data up to
  rank 11},'' \href{http://arxiv.org/abs/2308.09670}{{\ttfamily
  arXiv:2308.09670 [math.QA]}}.

\bibitem{Chang:2018iay}
C.-M. Chang, Y.-H. Lin, S.-H. Shao, Y.~Wang, and X.~Yin, ``{Topological Defect
  Lines and Renormalization Group Flows in Two Dimensions},''
  \href{http://dx.doi.org/10.1007/JHEP01(2019)026}{{\em JHEP} {\bfseries 01}
  (2019) 026}, \href{http://arxiv.org/abs/1802.04445}{{\ttfamily
  arXiv:1802.04445 [hep-th]}}.

\bibitem{Fendley2019}
E.~O'Brien, E.~Vernier, and P.~Fendley, ``{“Not-A”, representation
  symmetry-protected topological, and Potts phases in an S3-invariant chain},''
  {\em Physical Review B} {\bfseries 101} no.~23, (2020) 235108.

\bibitem{Hung:2025gcp}
L.-Y. Hung, K.~Ji, C.~Shen, Y.~Wan, and Y.~Zhao, ``{A 2D-CFT Factory: Critical
  Lattice Models from Competing Anyon Condensation Processes in
  SymTO/SymTFT},'' \href{http://arxiv.org/abs/2506.05324}{{\ttfamily
  arXiv:2506.05324 [cond-mat.str-el]}}.

\bibitem{Mong:2014ova}
R.~S.~K. Mong, D.~J. Clarke, J.~Alicea, N.~H. Lindner, and P.~Fendley,
  ``{Parafermionic conformal field theory on the lattice},''
  \href{http://dx.doi.org/10.1088/1751-8113/47/45/452001}{{\em J. Phys. A}
  {\bfseries 47} no.~45, (2014) 452001},
  \href{http://arxiv.org/abs/1406.0846}{{\ttfamily arXiv:1406.0846
  [cond-mat.stat-mech]}}.

\end{thebibliography}\endgroup

\newpage
\clearpage

\widetext
\begin{center}
\textbf{\large Supplementary Material: A Gapless Phase with Haagerup Symmetry}\\

\bigskip

{Lea E.\ Bottini and Sakura Sch\"afer-Nameki}\\
\smallskip

{\it Mathematical Institute, University
of Oxford, Woodstock Road, Oxford, OX2 6GG, United Kingdom}

\end{center}
\setcounter{equation}{0}
\setcounter{figure}{0}
\setcounter{table}{0}
\setcounter{page}{1}
\makeatletter
\renewcommand{\theequation}{S\arabic{equation}}
\renewcommand{\thefigure}{S\arabic{figure}}

\twocolumngrid


\section{Haagerup Symmetry $\cH_3$, Center and Algebras}

\subsection{Background on $\cH_3$}

We provide some background on the Haagerup $\cH_3$ fusion category, which will be important in carrying out the analysis, but is largely of technical nature. 
Starting with the simple objects
\begin{equation}
    \{ 1, \alpha, \alpha^2, \rho, \alpha \rho, \alpha^2 \rho \} \,,
\end{equation}
the first property is their quantum dimensions, which are 
\begin{equation}
    d_1 = d_\alpha = d_{\alpha^2} = 1 \quad , \quad d_\rho = d_{\alpha \rho} = d_{\alpha^2 \rho} = d \,,
\end{equation}
with $d= \frac{1}{2} \left(3+\sqrt{13}\right)$. The F-symbols for this fusion category are discussed in \cite{Huang:2020lox,osborne2019f}. We remark that $\cH_3$ is not one of the Haagerup fusion categories immediately descending from the Haagerup subfactor \cite{haagerup1994principal,asaeda1999exotic}, which are denoted $\cH_1$ and $\cH_2$, but is Morita equivalent to those, and was discovered in \cite{grossman2012quantum}. In particular, the fusion category $\cH_1$ has four simple objects, all non-invertible, and can be obtained from $\cH_3$ by gauging the algebra object $\cA_1 = 1 \oplus \rho \oplus \alpha 
\rho$, while $\cH_2$ has the same simple objects and fusion ring as $\cH_3$ and can be obtained from it by gauging the invertible $\Z_3$ sub-symmetry via the algebra object $\cA_2 = 1\oplus \alpha \oplus \alpha^2$. 

Essential for the analysis of phases is the Drinfeld center $\cZ(\cH_3)$, which was determined in \cite{Izumi:2001mi, hong2008exotic}. It has twelve simple objects
whose properties are summarized in  (\ref{eq:qdims_center}).

\be\label{eq:qdims_center}
\ba
&
\begin{array}{|c||c|c|c|c|c|c|}
\hline
\text{Obj} &
\id& \pi_1& \pi_2& \sigma_1& \sigma_2& \sigma_3\cr \hline
\text{Dim} &
1&3 d+1&3 d+2&3 d+2&3 d+2&3 d+2 
\cr \hline
\text{Spin} &1 & 1& 1& 1&  e^{\frac{2 i \pi }{3}}&e^{-\frac{2 i \pi }{3}} \cr \hline
\end{array}\cr 
&
\begin{array}{|c||c|c|c|c|c|c|}
\hline
\text{Obj} & \mu_1& \mu_2& \mu_3& \mu_4& \mu_5& \mu_6 \cr \hline
\text{Dim} & 3 d&3 d&3 d&3 d&3 d&3 d 
\cr \hline
\text{Spin} & e^{\frac{4 i \pi   }{13}}&e^{-\frac{4 i \pi }{13}}&e^{\frac{10 i \pi }{13}}&e^{-\frac{10 i \pi   }{13}}&e^{\frac{12 i \pi }{13}}&e^{-\frac{12 i \pi }{13}} \cr \hline
\end{array}
\ea
\ee


Notice there are three non-trivial bosons $ \sigma_1 \,, \pi_1$ and $\pi_2$. 
The $S$-matrix is 
\be
S= {1\over D}\left( \begin{array}{cc} 
A & B \\
B^t & C 
\end{array}
\right)\,,
\ee
where 
\be
D= \frac{3}{2} \left(13+3 \sqrt{13}\right) = \dim (\cZ (\cH))
\ee
is the total quantum dimension of $\cZ(\cH_3)$, 
\be
\footnotesize{
A= \left(\begin{array}{cccccc}
1 & 3 d+1 & 3 d+2 & 3 d+2 & 3 d+2 & 3 d+2 \\
3 d+1 & 1 & 3 d+2 & 3 d+2 & 3 d+2 & 3 d+2 \\
3 d+2 & 3 d+2 & 6 d+4 & -3 d-2 & -3 d-2 & -3 d-2 \\
3 d+2 & 3 d+2 & -3 d-2 & 6 d+4 & -3 d-2 & -3 d-2 \\
3 d+2 & 3 d+2 & -3 d-2 & -3 d-2 & -3 d-2 & 6 d+4 \\
3 d+2 & 3 d+2 & -3 d-2 & -3 d-2 & 6 d+4 & -3 d-2
\end{array}\right)
}
\ee
\be
B= 3 d\left(\begin{array}{cccccc}
1 & 1 & 1 & 1 & 1 & 1 \\
-1 & -1 & -1 & -1 & -1 & -1 \\
0 & 0 & 0 & 0 & 0 & 0 \\
0 & 0 & 0 & 0 & 0 & 0 \\
0 & 0 & 0 & 0 & 0 & 0 \\
0 & 0 & 0 & 0 & 0 & 0
\end{array}\right)
\ee
\be
C= 
3 d\left(\begin{array}{llllll}
x_1 & x_3 & x_6 & x_2 & x_4 & x_5 \\
x_3 & x_1 & x_2 & x_6 & x_5 & x_4 \\
x_6 & x_2 & x_4 & x_5 & x_1 & x_3 \\
x_2 & x_6 & x_5 & x_4 & x_3 & x_1 \\
x_4 & x_5 & x_1 & x_3 & x_6 & x_2 \\
x_5 & x_4 & x_3 & x_1 & x_2 & x_6
\end{array}\right)
\ee
and 
\be
\begin{aligned}
&\{x_1, \cdots, x_6\} \sim \\
&\{{0.7092,1.4970,1.9419,-0.2411,-1.1361,-1.7709}\}
\end{aligned}
\ee
are the roots of 
\be
P(x)= x^6-x^5-5 x^4+4 x^3+6 x^2-3 x-1 \,.
\ee

Let us recall the possible representations of the Haagerup fusion ring, following \cite{Huang:2021ytb}.
 Since the two generators $\rho$ and $\alpha$ do not commute, there is no basis of local operators in which their action can be simultaneously diagonalised. If we choose a basis that diagonalises the $\Z_3$ action, the irreducible representations of $\cH_3$ are: 
\begin{itemize}
    \item[-] a 1-dimensional representation $1^+$, where $\alpha$ acts trivially and $\rho$ acts as $d = \frac{3+\sqrt{13}}{2}$;
    \item[-] a 1-dimensional representation $1^-$, where $\alpha$ acts trivially and $\rho$ acts as $-d^{-1} = \frac{3-\sqrt{13}}{2}$;
    \item[-] a 2-dimensional representation $\bm{2}$, where $\alpha$ acts as $\begin{pmatrix}
        \omega &  0\\
        0 & \omega^2 
    \end{pmatrix}$ and $\rho$ acts as $\begin{pmatrix}
        0 & 1 \\ 1 & 0
    \end{pmatrix}$ \,,
\end{itemize}
where $\omega = e^{2\pi i / 3}$.
We remark that in the above by `action' we mean the linking action of $\rho$ and $\alpha$ on local operators, obtained by encircling a local operator with a line and shrinking it to zero size to obtain a new local operator. One could consider also a `passing through' action, which in the case of the non-invertible generator $\rho$ involves mixing together local and twisted sector operators into a single multiplet $\cM$. The possible generalised charges under $\cH_3$, i.e.\ multiplets $\cM$, are in one-to-one correspondence with anyons in the Drinfeld center $\cZ(\cH_3)$ \cite{Lin:2022dhv,Bhardwaj:2023ayw,Bartsch:2023wvv}.

\subsection{Condensable Algebras}\label{app:algebras}

The general discussion of the necessary conditions that Lagrangian and condensable algebras need to satisfy can be found e.g.\ in \cite{Ng:2023wsc}.
The Lagrangian algebras are
\be
\ba
\cL_1 &= 1 \oplus \pi_1 \oplus 2 \sigma_1 \\
\cL_2 &=1 \oplus \pi_1 \oplus \pi_2 \oplus \sigma_1\\
\cL_3 &= 1  \oplus \pi_1 \oplus 2 \pi_ 2  \,.
\ea
\ee
We find that the only condensable non-Lagrangian algebras in $\cZ(\cH_3)$ are 
\be
\ba
\cA_0 &= 1 \cr 
\cA_1&  = 1 \oplus \pi_1  \,. \cr
\ea
\ee
Notice that the identity line satisfies in a non-trivial way the cyclotomic integer condition, as the ratio of S-matrices contains the following distinct entries
\be
\left\{1,\frac{1}{2} \left(11+3 \sqrt{13}\right) ,\frac{1}{2} \left(13+3 \sqrt{13}\right),\frac{3}{2}
   \left(3+\sqrt{13}\right)\right\} \,,
\ee
which all turn out to be cyclotomic integers in a non-trivial way. 
For instance,
\begin{equation}
\begin{aligned}
&\frac{1}{2} \left(11+3 \sqrt{13}\right) = 
-4E_{13}-7E_{13}^2-4E_{13}^3-4E_{13}^4-7E_{13}^5 + \\
&-7E_{13}^6-7E_{13}^7-7E_{13}^8-4E_{13}^9-4E_{13}^{10}-7E_{13}^{11} -4E_{13}^{12} \,,
\end{aligned}
\end{equation}
where $E_{13}= e^{ 2\pi i/13}$. We checked these conditions using GAP.

The other intersections between the Lagrangian algebras, namely $1 \oplus \pi_1 \oplus \sigma_1$ and $1 \oplus \pi_1 \oplus \pi_2$, are not condensable. The reason they are not condensable is that the cyclotomic integer condition is not satisfied, as the S-matrix ratios include e.g.\ $1/2$, and all rational numbers are not cyclotomic integers.  Finally, $1\oplus \pi_2$ and $1\oplus \sigma_1$ are not condensable as the S-matrix ratio takes the value $1/6 (7+\sqrt{13})$, which again is not a cyclotomic integer.

\section{Gapped Phases with Haagerup Symmetry}\label{app:gapped}

In this section we provide more details regarding the derivation of the three gapped phases with Haagerup symmetry using the SymTFT construction outlined in the main text. In particular, we determine the  order parameters and derive the action of the symmetry on the gapped phases. 

\paragraph{\textbf{2 Vacua Phase} :}
This corresponds to the choice $\cL_\phys = \cL_1$ for the physical boundary.
As only the identity line and the $\pi_1$ line can end on both boundaries, the resulting TQFT has two vacua. This is succinctly summarised in the SymTFT picture:
\begin{equation}
\begin{tikzpicture}
\begin{scope}[shift={(0,0)}]
\draw [fill= black, opacity =0.1] (0,-1) -- (0,1)--(2,1) -- (2, -1) -- (0,-1) ; 
\draw [thick] (0,-1) -- (0,1);
\draw [thick] (2,-1) -- (2,1);
\draw [thick] (0,0.4) -- (2,0.4) ;
\draw [thick] (0, -0.4) -- (2, -0.4) ;
\node[above] at (1,0.5) {$1$} ;
\node[above ] at (1,-0.35) {$\pi_1$} ;
\node[above] at (0,1) {$\cL_3$}; 
\node[above] at (2,1) {$\cL_1$}; 
\node at (2.5,0.4) {$1$};
\node at (2.5,-0.4) {$\cO_1$};
\draw [black,fill=black] (0,-0.4) ellipse (0.05 and 0.05);
\draw [black,fill=black] (2,-0.4) ellipse (0.05 and 0.05);
\draw [black,fill=black] (0,0.4) ellipse (0.05 and 0.05);
\draw [black,fill=black] (2,0.4) ellipse (0.05 and 0.05);
\end{scope}
\end{tikzpicture}
\end{equation}
The resulting local operators naturally transform in representations of the Haagerup fusion ring. In particular $O_1$, being a single local operator, must transform in the non-trivial 1-dimensional representation $1^-$. This means that the linking action of the $\cH_3$ generators on the local operators is the following 
\begin{equation}\label{eq:2vacua_H3action}
\begin{aligned}
    &\alpha: 1\rightarrow 1 \quad,\quad \cO_1 \rightarrow \cO_1 \\
   &\rho: 1 \rightarrow d \quad,\quad \cO_1 \rightarrow -d^{-1} \cO_1 \,.
    \end{aligned}
\end{equation}
The fusion of the bulk line $\pi_1$ with itself is 
\begin{equation}
    \pi_1 \otimes \pi_1 = 1 \oplus \pi_1 \oplus \pi_2 \oplus \bigoplus_{i=1}^3 \sigma_i \oplus \bigoplus_{i=1}^6 \mu_i \,.
\end{equation}
This, together with the fact that $\cO_1$ is neutral under the invertible $\Z_3$ subsymmetry of $\cH_3$, constrains the product of $\cO_1$ with itself to be 
\begin{equation}
    \cO_1 \cO_1 = 1 + \lambda \cO_1 \,. 
\end{equation}
We can use this to determine the two vacua $v_i$, $i=1,2$, satisfying 
\begin{equation}
    v_i v_j = \delta_{ij} v_j
\end{equation}
to be
\begin{equation}
\begin{aligned}
    v_0 &= \frac{\sqrt{4+\lambda^2} - \lambda}{2\sqrt{4+\lambda^2}} + \frac{\cO_1}{\sqrt{4+\lambda^2}} \\ 
    v_1 &= \frac{\sqrt{4+\lambda^2} + \lambda}{2\sqrt{4+\lambda^2}} - \frac{\cO_1}{\sqrt{4+\lambda^2}} \,.
\end{aligned}    
\end{equation}
Now let us consider the $\cH_3$ symmetry action on the vacua. Clearly, since $\cO_1$ is uncharged under $\Z_3$, the action of the generator $\alpha$ on the vacua is trivial. Therefore, we can represent it on the vacua as 
\begin{equation}\label{eq:2vacua_alpha}
    \alpha = \alpha^2 = 1 = 1_{00} \oplus 1_{11} \,.
\end{equation}
Here and in the following this notation $1_{ab}$  means an interface from  $a$ to $b$, namely a map $1_{ab} v_c = \zeta_{ab} \delta_{ac} v_b$, where $\zeta_{ab}$ are possible Euler counter-terms.
The action of the non-invertible generator $\rho$ can be determined using  \eqref{eq:2vacua_H3action} to be
\begin{equation}
\begin{aligned}
    v_0 \rightarrow d \, \frac{\sqrt{4+\lambda^2} - \lambda}{2\sqrt{4+\lambda^2}} - d^{-1} \frac{\cO_1}{\sqrt{4+\lambda^2}} = A v_0 + B v_1 \\
    v_1 \rightarrow d\, \frac{\sqrt{4+\lambda^2} + \lambda}{2\sqrt{4+\lambda^2}} + d^{-1} \frac{\cO_1}{\sqrt{4+\lambda^2}} = C v_0 + D v_1 \,,
\end{aligned}
\end{equation}
where the coefficients are given by 
\begin{equation}
\begin{aligned}
    A &= \frac{6(3+\sqrt{13})-2\lambda (\lambda+\sqrt{4+\lambda^2})}{(3+\sqrt{13})(4+\lambda(\lambda+\sqrt{4+\lambda^2}))}\\ 
    B &= \frac{2 \sqrt{13}}{4+\lambda(\lambda+\sqrt{4+\lambda^2})}\\
    C &= \frac{\sqrt{13}(\lambda+\sqrt{4+\lambda^2})}{2\sqrt{4+\lambda^2}} \\
    D &= \frac{3}{2} + \frac{\sqrt{13}\lambda}{2\sqrt{4+\lambda^2}} \,.
\end{aligned}    
\end{equation}
Notice in particular that we have 
\begin{equation}
    1_{01} : v_0 \rightarrow B v_1 \quad,\quad 1_{10} : v_1 \rightarrow C v_0 \,,
\end{equation}
where $B$ and $C$ encode the non-trivial Euler counter-terms. A consistent action requires $B C = 1$. This fixes $\lambda$ to be one of the two possible values $\lambda = \pm 3$. To pin down the sign, we can for example require that the symmetry action on $v_0$ produces an integer value of $v_0$, i.e.\ we demand $A$ to be integer, as a non-integer value would not allow us to represent $\rho$ by any line operator in this phase. This fixes $\lambda=3$. In summary, we have found $\cO_1 \cO_1 = 1+3\cO_1$, with vacua 
\begin{equation}
\begin{aligned}
    &v_0 = \frac{d^{-1}+ \cO_1}{\sqrt{13}} \\
    &v_1 = \frac{d-\cO_1}{\sqrt{13}} \,.
\end{aligned}    
\end{equation}
With $\lambda=3$, the symmetry action of $\rho$ becomes 
\begin{equation}
    \rho: v_0 \rightarrow d^{-1}\, v_1 \quad,\quad v_1 \rightarrow d \,v_0    + 3 v_1 \,. 
\end{equation}
Hence we can represent $\rho$ as 
\begin{equation}\label{eq:2vacua_rho}
    \rho =  1_{01} \oplus 1_{10} \oplus 3 \,1_{11} \,.
\end{equation}
One can check that 
\begin{equation}
\begin{aligned}
    \rho \otimes \rho &= 1_{00} \oplus 1_{11} \oplus 3 (1_{01} \oplus 1_{10} \oplus 3\, 1_{11}) \\
    &= 1 \oplus \rho \oplus \alpha \rho \oplus \alpha^2 \rho \,.
\end{aligned}    
\end{equation}
This gapped phase can be characterised in terms of the vev of a local operator $\cO_1$ transforming in the $1^-$ representation of $\cH_3$. This local operator $\cO_1$ belongs to a multiplet of the $\cH_3$ action labelled by the anyon $\pi_1$ which contains also $\alpha^i \rho$-twisted sector operators, $i=0,1,2$. This phase is moreover characterised by the condensation of two purely string order parameters which sit in a multiplet labelled by the bulk anyon $\sigma_1$, comprising of operators in the $\Z_3$ subsymmetry twisted sector and $\alpha^i \rho$-twisted sector, $i=0,1,2$. 

\paragraph{\textbf{4 Vacua Phase:}}

We now move to consider the choice $\cL_\phys = \cL_2$ for the physical boundary.
In this case the identity $\pi_1$ and $\pi_2$ lines can end on both boundaries
\begin{equation}
\begin{tikzpicture}
\begin{scope}[shift={(0,0)}]
\draw [fill= black, opacity =0.1] (0,-2) -- (0,0.5)--(2,0.5) -- (2, -2) -- (0,-2) ; 
\draw [thick] (0,-2) -- (0,0.65) ;
\draw [thick] (2,-2) -- (2,0.65) ;
\draw [thick] (0,0) -- (2,0) ;
\draw [thick] (0, -0.75) -- (2, -0.75) ;
\draw [thick] (0, -1.5) -- (2, -1.5) ;
\node[above] at (1,0) {$\pi_2$} ;
\node[above ] at (1,-0.75) {$\pi_1$} ;
\node[above ] at (1,-1.5) {$1$} ;
\node[above] at (0,0.65) {$\cL_3$}; 
\node[above] at (2,0.65) {$\cL_2$}; 
\node at (2.5,0) {$\cO_2^i$};
\node at (2.5,-0.75) {$\cO_1$};
\node at (2.5,-1.5) {$1$};
\draw [black,fill=black] (0,-1.5) ellipse (0.05 and 0.05);
\draw [black,fill=black] (0,-0.75) ellipse (0.05 and 0.05);
\draw [black,fill=black] (0,0) ellipse (0.05 and 0.05);
\draw [black,fill=black] (2,-0.75) ellipse (0.05 and 0.05);
\draw [black,fill=black] (2,-1.5) ellipse (0.05 and 0.05);
\draw [black,fill=black] (2,0) ellipse (0.05 and 0.05);
\end{scope}
\end{tikzpicture}
\end{equation}
After compactification, we get four local operator $1$, $\cO_1$ and $\cO_2^i$, $i=1,2$, which follows from the fact that the $\pi_2$ line appears with multiplicity two in the reference Lagrangian algebra $\cL_3$. These operators naturally transform in representations of the Haagerup fusion ring. In particular, $\cO_1$ transforms in the $1^-$ representation, while the doublet $\cO_2^i$ transforms in the $\bm{2}$ representation. It follows that the linking action of the Haagerup symmetry generators $\alpha$ and $\rho$ on the local operators is 
\begin{equation}\label{eq:4vacua_H3action}
\begin{aligned}
   &\alpha: \cO_1 \rightarrow \cO_1 \quad,\quad \cO_2^i \rightarrow \omega^i \cO_2^i \\
   &\rho: \cO_1 \rightarrow -d^{-1} \cO_1 \quad,\quad \cO_2^1 \leftrightarrow \cO_2^2 \,.
\end{aligned}   
\end{equation}
Using the bulk lines fusions 
\begin{equation}
\begin{aligned}
    &\pi_1 \otimes \pi_1 = 1 \oplus \pi_1 \oplus \pi_2 \oplus \bigoplus_{i=1}^3 \sigma_i \oplus \bigoplus_{i=1}^6 \mu_i \\
    &\pi_2 \otimes \pi_2 = 1 \oplus 2 \pi_1 \oplus 2 \pi_2 \oplus \bigoplus_{i=1}^3 \sigma_i  \oplus \bigoplus_{i=1}^6 \mu_i \\ 
    &\pi_1 \otimes \pi_2 = \pi_1 \oplus 2\pi_2 \oplus \bigoplus_{i=1}^3 \sigma_i  \oplus \bigoplus_{i=1}^6 \mu_i \,.
\end{aligned}    
\end{equation}
and the $\Z_3$ symmetry action, we can constrain the product of the local operators to be 
\begin{equation}
\begin{aligned}
    &\cO_1 \cO_1 = 1 + \lambda \cO_1 \,,\quad \cO_1 \cO_2^1 = \beta_1 \cO_2^1 \,,\quad \cO_1 \cO_2^2 = \beta_2 \cO_2^2 \\
    &\cO_2^1 \cO_2^1 = \gamma \cO_2^2 \,,\quad \cO_2^2 \cO_2^2 = \gamma \cO_2^1 \,,\quad \cO_2^1 \cO_2^2 = 1 + \delta \cO_1 \,.
\end{aligned}    
\end{equation}
Imposing associativity for the product $\cO_1 \cO_1 \cO_2^i$ moreover fixes 
$
    \beta_1 = \frac{\lambda \pm \sqrt{4+\lambda^2}}{2},$ $\beta_2 = \frac{\lambda \pm \sqrt{4+\lambda^2}}{2}
$.
Associativity for the product $\cO_1 \cO_2^i \cO_2^i$ fixes $\beta_1 = \beta_2=\beta$, and associativity for the product $\cO_1 \cO_2^1 \cO_2^2$ fixes $\delta = \beta$. Finally, associativity for the product $\cO_2^1 \cO_2^1 \cO_2^2$ fixes $\gamma = \sqrt{1+\beta^2}$. 
Notice that every parameter appearing in the operator products is fixed in terms of $\lambda$ and that we find two possible TQFTs depending on the choice for $\beta$. Without loss of generality, we choose the positive root. The four vacua turn out to be 
\begin{equation}
\begin{aligned}
    &v_0 = \frac{\beta^{-1}+\cO_1+(4+\lambda^2)^{1/4}\sqrt{\beta^{-1}}( \cO_2^1 + \cO_2^2)}{3\sqrt{4+\lambda^2}} \\ 
    &v_1 = \frac{\beta^{-1}+\cO_1+(4+\lambda^2)^{1/4}\sqrt{\beta^{-1}} (\omega \cO_2^1 + \omega^2 \cO_2^2)}{3\sqrt{4+\lambda^2}} \\ 
    &v_2 = \frac{\beta^{-1}+\cO_1+(4+\lambda^2)^{1/4}\sqrt{\beta^{-1}} (\omega^2 \cO_2^1 + \omega \cO_2^2)}{3\sqrt{4+\lambda^2}}  \\ 
    &v_3 = \frac{\beta-\cO_1}{\sqrt{4+\lambda^2}}  \,.
\end{aligned}    
\end{equation}

Now let us consider the $\cH_3$ symmetry action on the vacua. The invertible $\Z_3$ symmetry generator $\alpha$ clearly permutes the first three vacua and leaves $v_3$ fixed, and therefore we can identify 
\begin{equation}\label{eq:alpha_4vacua}
    \alpha = 1_{01} \oplus 1_{12} \oplus 1_{20} \oplus 1_{33} \,.
\end{equation}
Notice that from the point of view of the invertible $\Z_3$ symmetry, this phase decomposes as a the sum of a $\Z_3$ SSB phase and a $\Z_3$ trivial phase.
The non-invertible generator $\rho$ acts in a more complicated manner, following \eqref{eq:4vacua_H3action}. As before, we would like to determine the coefficient $\lambda$ by having a consistent action of the non-invertible symmetry on the vacua once Euler counter-terms are taken into account. We start by considering the action of $\rho$ on $v_3$. This is 
\begin{equation}
    \rho: v_3 \rightarrow A (v_0 + v_1 + v_2) + B v_3 \,,
\end{equation}
where the coefficients are 
\begin{equation}
    A = \frac{\sqrt{13}(\lambda+\sqrt{4+\lambda^2})}{2\sqrt{4+\lambda^2}} \quad,\quad B = \frac{3}{2}+ \frac{\sqrt{13}\lambda}{2\sqrt{4+\lambda^2}} \,.
\end{equation}
Now consider the action of $\rho$ on $v_1$, which we compute to be 
\begin{equation}
    v_0 \rightarrow C v_0 + D v_1 + E v_2 + F v_3 \,,
\end{equation}
with coefficients 
\begin{equation}
\begin{aligned}
    &C = \frac{7}{6} - \frac{\sqrt{13}\lambda}{6\sqrt{4+\lambda^2}} \\
    &D=E = \frac{1}{6} - \frac{\sqrt{13}\lambda}{6\sqrt{4+\lambda^2}}  \\
    &F = \frac{2\sqrt{13}}{3(4+\lambda(\lambda+\sqrt{4+\lambda^2}))} 
    \end{aligned}
\end{equation}
We start by imposing $AF=1$ to have a consistent action $v_3 \rightarrow v_0 \rightarrow v_3$. This gives the two options $\lambda = \pm 1/\sqrt{3}$. We can pin down the sign by requiring for example that $C$ is integer. This gives the solution 
\begin{equation}
    \lambda = \frac{1}{\sqrt{3}} \,.
\end{equation}
For this specific value, the action of $\rho$ on all the vacua becomes the following 
\begin{equation}
\begin{aligned}
    \rho: \qquad &v_0 \rightarrow v_0 + a^{-1} v_3 \\
    &v_1 \rightarrow v_2 + a^{-1} v_3 \\
    &v_2 \rightarrow v_1 + a^{-1} v_3 \\
    &v_3 \rightarrow a (v_0+v_1+v_2) + 2 v_3 \,.
\end{aligned}    
\end{equation}
with $a = \frac{1+\sqrt{13}}{2}$. In summary, we can identify $\rho$ as 
\begin{equation}\label{eq:rho_4vacua}
    \rho = 1_{00} \oplus  1_{03} \oplus 1_{12} \oplus 1_{13} \oplus 1_{21} \oplus  1_{23} \oplus 1_{30} \oplus 1_{31} \oplus 1_{32} \oplus 2\,  1_{33} \,.
\end{equation}
We can easily obtain $\alpha \rho$ and $\alpha^2 \rho$ by composing \eqref{eq:alpha_4vacua} with \eqref{eq:rho_4vacua}. Then it is easily checked that our expression for $\rho$ satisfies the fusion $\rho^2 = 1 \oplus \rho \oplus \alpha \rho \oplus \alpha^2 \rho$.

This phase can be characterised in terms of the condensation of a local operator $\cO_1$ transforming in the $1^-$ representations of Haagerup and local operators $\cO_2^1$ and $\cO_2^1$ transforming as a doublet of the $\bm{2}$ representation of $\cH_3$. $\cO_1$ sits again in a $\cH_3$ multiplet labeled by the anyon $\pi_1$, while $\cO_2^i$ sit in a $\cH_3$ multiplet labeled by the anyon $\pi_2$. Both of these also involve $\alpha^i \rho$-twisted operators. Finally, this phase also has a string order parameter labeled by the anyon $\sigma_1$. 
\paragraph{\textbf{6 Vacua Phase:}}
A gapped phases where the full $\cH_3$ symmetry is spontaneously broken is obtained by selecting $\cL_\phys = \cL_3$.
The resulting TQFT has 6 vacua, as follows from the fact that the $\pi_1$ and $\pi_2$ lines can terminate on both boundaries, the latter with multiplicity two 
\begin{equation}
\begin{tikzpicture}
\begin{scope}[shift={(0,0)}]
\draw [fill= black, opacity =0.1] (0,-2) -- (0,1)--(2,1) -- (2, -2) -- (0,-2) ; 
\draw [thick] (0,-2) -- (0,1) ;
\draw [thick] (2,-2) -- (2,1) ;
\draw [thick] (0,0.5) -- (2,0.5) ;
\draw [thick] (0,0) -- (2,0) ;
\draw [thick] (0, -0.75) -- (2, -0.75) ;
\draw [thick] (0, -1.5) -- (2, -1.5) ;
\node[above] at (1,0.5) {$\pi_2$} ;
\node[above ] at (1,-0.75) {$\pi_1$} ;
\node[above ] at (1,-1.5) {$1$} ;
\node[above] at (0,1) {$\cL_3$}; 
\node[above] at (2,1) {$\cL_3$}; 
\draw [thick] (0,0.5) -- (2,0) ;
\draw [thick] (0,0) -- (2,0.5) ;
\draw [black,fill=black] (0,0.5) ellipse (0.05 and 0.05);
\node at (2.5,0.25) {$\cO_2^{i,j}$};
\node at (2.5,-0.75) {$\cO_1$};
\node at (2.5,-1.5) {$1$};
\draw [black,fill=black] (0,-1.5) ellipse (0.05 and 0.05);
\draw [black,fill=black] (0,-0.75) ellipse (0.05 and 0.05);
\draw [black,fill=black] (0,0) ellipse (0.05 and 0.05);
\draw [black,fill=black] (2,0.5) ellipse (0.05 and 0.05);
\draw [black,fill=black] (2,-0.75) ellipse (0.05 and 0.05);
\draw [black,fill=black] (2,-1.5) ellipse (0.05 and 0.05);
\draw [black,fill=black] (2,0) ellipse (0.05 and 0.05);
\end{scope}
\end{tikzpicture}
\end{equation}
After compactification, we get six local operators. We denote by $\cO_1$ the non-trivial operator coming from $\pi_1$ and by $\cO_2^{i,j}$ the four non-trivial operators coming from $\pi_2$.
These operators naturally transform in representations of the Haagerup symmetry.
The local operator $1$, coming from the identity line, naturally transforms in the $1^+$ representation, while  $\cO_1$ and $\cO_2^{i,j}$, coming from $\pi_1$ and $\pi_2$ respectively, transform in the $1^-$ and $\bm{2}$ representations. In summary, the (linking) action of the left and right generators of $\cH_3$ on the local operators resulting from the compactification is 
\begin{equation}
\begin{aligned}
    &\alpha_L: 1 \rightarrow 1 \qquad \cO_1 \rightarrow \cO_1 \qquad \cO_2^{i,j} \rightarrow \omega^i \, \cO_2^{i,j} \\ 
     &\alpha_R: 1 \rightarrow 1 \qquad \cO_1 \rightarrow \cO_1 \qquad \cO_2^{i,j} \rightarrow \omega^j \, \cO_2^{i,j} 
\end{aligned}    
\end{equation}
and 
\begin{equation}\label{eq:rho_6vacua}
\begin{aligned}
    &\rho_L: 1 \rightarrow d  \qquad \cO_1 \rightarrow -d^{-1} \cO_1 \qquad  \cO_2^{1,j} \leftrightarrow \cO_2^{2,j} \\ 
    &\rho_R: 1 \rightarrow d  \qquad \cO_1 \rightarrow -d^{-1} \cO_1 \qquad  \cO_2^{i,1} \leftrightarrow \cO_2^{i,2} \,.
\end{aligned}    
\end{equation}
We can use both the left and right symmetry action and the fusion of the bulk lines to constrain the product of the local operators. 
These fix the products to be 
\begin{equation}
\begin{aligned}
    &\cO_1^2 = 1 + \lambda \cO_1 \,, \quad \cO_1 \cO_2^{i,i} = \tilde{\beta} \cO_2^{i,i} \,, \quad \cO_1 \cO_2^{i,j} = \beta \cO_2^{i,j} \\
    &(\cO_2^{i,i})^2 = \gamma \cO_2^{i+1,i+1} \,, \quad (\cO_2^{i,j})^2 = \delta \cO_2^{i+1,j+1} \\
    &\cO_2^{1,2} \cO_2^{2,1} = 1 + \epsilon \cO_1 \,, \quad
    \cO_2^{1,1} \cO_2^{2,2} = 1 + \rho \cO_1 \\
    &\cO_2^{i,i} \cO_2^{i+1,i} = \cO_2^{i,i} \cO_2^{i,i+1} = 0\,, \quad i\neq j \,, i,j=1,2\,.
\end{aligned}    
\end{equation}
Now we can impose various associativity relations. For example, associativity for the product $\cO_1 \cO_2^{i,j} \cO_2^{i,j}$ gives for $\beta$ the two possible values
$\beta = 1/2(\lambda \pm \sqrt{4+\lambda^2})$. Without loss of generality, we choose the positive root. We can then fix for $\tilde{\beta}$ the negative root, $\tilde{\beta} = - \beta^{-1}$. 
Associativity for the product $\cO_1 \cO_2^{1,1} \cO_2^{2,2}$ fixes then $\rho = -\beta^{-1}$. Similarly, we have $\epsilon = \beta$. Finally, associativity for the product $\cO_2^{1,1} \cO_2^{2,2} \cO_2^{2,2}$ fixes $\gamma = \sqrt{1+\beta^{-2}}$, and analogously we fix $\delta = \sqrt{1+\beta^{2}}$. In summary, all the product of local operators are fixed in terms on the coefficient $\lambda$ appearing in the product of $\cO_1$ with itself. 

Having this, we can determine the 6 vacua of the TQFT, which are given by 
\begin{equation}
\begin{aligned}
    &v_0 = \frac{\beta- \cO_1 + (4+\lambda^2)^{1/4}\sqrt{\beta}(\cO_2^{1,1}+\cO_2^{2,2})}{3\sqrt{4+\lambda^2}}  \\ 
     &v_1 = \frac{\beta-\cO_1+(4+\lambda^2)^{1/4}\sqrt{\beta}(\omega \cO_2^{1,1}+\omega^2 \cO_2^{2,2})}{3\sqrt{4+\lambda^2}} \\ 
     &v_2 = \frac{\beta-\cO_1+(4+\lambda^2)^{1/4}\sqrt{\beta}(\omega^2 \cO_2^{1,1}+\omega \cO_2^{2,2})}{3\sqrt{4+\lambda^2}} \\ 
      &v_3 = \frac{\beta^{-1}+\cO_1+(4+\lambda^2)^{1/4}\sqrt{\beta^{-1}}(\cO_2^{1,2}+\cO_2^{2,1})}{3\sqrt{4+\lambda^2}}  \\ 
      &v_4 = \frac{\beta^{-1}+\cO_1+(4+\lambda^2)^{1/4}\sqrt{\beta^{-1}}(\omega \cO_2^{1,2}+ \omega^2 \cO_2^{2,1})}{3\sqrt{4+\lambda^2}} \\
      &v_5 = \frac{\beta^{-1}+\cO_1+(4+\lambda^2)^{1/4}\sqrt{\beta^{-1}}(\omega^2 \cO_2^{1,2}+\omega \cO_2^{2,1})}{3\sqrt{4+\lambda^2}} \,.
\end{aligned}    
\end{equation}

Now let us consider the $\cH_3$ symmetry action of the vacua. We start with the $\Z_3$ invertible symmetry, which permutes the first set of vacua $v_0$, $v_1$ and $v_2$ among themselves, as well as the second set of vacua $v_3$, $v_4$ and $v_5$. We can therefore identify 
\begin{equation}\label{eq:alpha_6vacua}
    \alpha = 1_{01} \oplus 1_{12} \oplus 1_{20} \oplus 1_{34} \oplus 1_{45} \oplus 1 _{53} \,.
\end{equation}
We can also compute the action of the non-invertible generator $\rho$ on the vacua using \eqref{eq:rho_6vacua}. Using the same constraints as before, namely that the coefficient of a $1_{ii}$ term must be integer and that the coefficients of two terms $1_{ij}$ and $1_{ji}$ must multiply to one, we can determine 
\begin{equation}
    \lambda = 3 \,.
\end{equation}
We write down the symmetry action only for this consistent value to avoid cluttered expressions. Having fixed $\alpha$, the $\rho$ action is given by 
\begin{equation}
\begin{aligned}
    \rho: \qquad &v_0 \rightarrow v_0 + v_1 + v_2 + d \,v_3 \\ 
    &v_1 \rightarrow v_0 + v_1 + v_2 + d \,v_5 \\ 
    &v_2 \rightarrow v_0 + v_1 + v_2 + d \,v_4 \\ 
    &v_3 \rightarrow d^{-1} v_0  \\ 
    &v_4 \rightarrow d^{-1} v_2  \\
    &v_5 \rightarrow d^{-1} v_1  \,.
\end{aligned}    
\end{equation}
Therefore we can identify 
\begin{equation}\label{eq:rho_6vacua}
\begin{aligned}
    \rho &= 1_{00} \oplus 1_{01} \oplus 1_{02} \oplus 1_{03} \oplus 1_{10} \oplus 1_{11} \oplus 1_{12} \oplus 1_{15} \oplus \\
    &\oplus 1_{20} \oplus 1_{21} \oplus 1_{22} \oplus 1_{24} \oplus
     1_{30} \oplus 1_{42} \oplus 1_{51} \,.
\end{aligned}
\end{equation}
As usual, one can easily check that this $\rho$ satisfies the proper fusion algebra.

This phase can be characterised in terms of the condensation of the local operators $\cO_1$, transforming in the $1^-$ representation, $\cO_2^{1,1}$ and $\cO_2^{2,1}$, transforming in the $\bm{2}$ representation, and another doublet $\cO_2^{1,2}$ and $\cO_2^{2,2}$ in the $\bm{2}$. There are no multiplets involving purely string order parameters condensing in this phase.

\section{Gapless Phases and Phase Diagram}

We now consider the condensation of the non-Lagrangian algebra $\cA_1 = 1 \oplus \pi_1$ in $\cZ(\cH_3)$. 

Recall that if we condense an algebra $\cA$, the anyons that survive the condensation are those that braid trivially with the ones appearing in the algebra. If an algebra is Lagrangian, i.e.\ maximal, all the anyons not in $\cA$ braid non-trivially with at least one anyon in $\cA$, and therefore the reduced topological order $\cZ(\cS) / \cA$ is trivial. Therefore, a Lagrangian algebra specifies a gapped interface between a topological order and the trivial theory, or in other words a gapped boundary. If an algebra is non-Lagrangian, some anyons still braid trivially with the anyons in $\cA$. In this case, the reduced topological order $\cZ(\cS) / \cA$ is non-trivial and the algebra $\cA$ actually specifies a gapped interface. We can moreover express $\cZ(\cS) / \cA$ as the Drinfeld center of a “smaller" symmetry $\cS'$ (note $\cS' = \cS$ if $\cA$ is trivial, $\cA = 1$, in which case the interface is trivial). 

In this context it is useful to define the 
club quiche  $\cQ_{\cS,\cS'}$, which consists of the addition of a gapped interface to the standard SymTFT quiche, i.e.\ it comprises of two topological orders $\cZ(\cS)$ and $\cZ(\cS')$ separated by an interface $\cI$ specified by a condensable algebra $\cA \in \cZ(\cS)$:
\be\label{eq:club_quiche}
\begin{tikzpicture}
\begin{scope}[shift={(0,0)}]
\node at (-1.5,1.3) {$\cQ_{\cS,\cS'}:$};
\draw [fill= black, opacity =0.1] (0,0.6) -- (0,2) -- (2,2) -- (2,0.6) -- (0,0.6)  ; 
\draw [fill= black, opacity =0.1] (4,0.6) -- (4,2) -- (2,2) -- (2,0.6) -- (4,0.6) ; 
\draw [very thick] (0,0.6) -- (0,2) ;
\draw [very thick] (2,0.6) -- (2,2) ;
\node at (1,1.3) {$\cZ(\cS)$} ;
\node at (3,1.3) {$\cZ(\cS')$} ;
\node[above] at (0,2) {$\Bsym_\cS$}; 
\node[above] at (2,2) {$\cI$}; 
\end{scope}
\end{tikzpicture}
\ee
The club-quiche construction equivalently determines an $\cS$-symmetric boundary condition of the reduced topological order (by collapsing the first interval).
Inputting a physical boundary condition on the RHS yields the club sandwich, which allows us to construct the $\cS$-symmetric theories, in particular phase transitions, starting from $\cS'$-symmetric ones, as explained in the main text.

For the Haagerup symmetry, we consider the algebra $\cA_1$: the total quantum dimension of $\cZ(\cH_3)$ is $3(3d+2)$ and the quantum dimension of $\cA_1$ is $(3d+2)$, so the quantum dimension of the reduced topological order is
\begin{equation}
    D(\cZ(\cH_3) / \cA_1) = D(\cZ(\cH_3)) / D(\cA_1) = 3\,.
\end{equation}
This, together with the fact that the anyons in $\cH_3$ braiding trivially with $\pi_1$ are
\begin{equation}
    \{ 1, \pi_1, \pi_2, \sigma_1, \sigma_2, \sigma_3 \}\,,
\end{equation}
which have topological spins respectively 
\begin{equation}
    \{ 1,1,1,1,e^{\frac{2\pi i}{3}},e^{-\frac{2\pi i}{3}}\} \,,
\end{equation}
leads us naturally to conjecture that 
the reduced topological order $\cZ'$ is the $\Z_3$ Dijkgraaf-Witten theory, with anyon content 
\begin{equation}
    \cZ' = \cZ(\Z_3) = \{1,e,e^2,m,m^2,em,e^2m,em^2,e^2m^2\} \,.
\end{equation}
Notice that using the folding technique we can re-express the algebra $\cA$ defining an interface between $\cZ(\cH_3)$ and $\cZ(\Z_3)$ as a Lagrangian algebra $\cL_\cA$ for the folded theory $\cZ(\cH_3)\boxtimes \overline{\cZ(\Z_3)}$, which in the present case is 
\begin{equation}
\begin{aligned}
    \cL &= 1 \oplus \pi_1 \oplus \pi_2 (\overline{e}\oplus \overline{e}^2) \oplus \sigma_1 (\overline{m} \oplus \overline{m}^2) \oplus \\
    &\oplus \sigma_2 (\overline{em}^2 \oplus \overline{e}^2\overline{m}) \oplus \sigma_3 (\overline{em} \oplus \overline{e}^2\overline{m}^2) \,.
\end{aligned}
\end{equation}
This is unique up to the automorphism of $\cZ(\Z_3)$ that exchanges $e$ and $m$. 

The club quiche picture we obtain is then the following: 
\begin{equation}
\label{CQA1}
\begin{tikzpicture}
\begin{scope}[shift={(0,0)}]
\draw [fill= black, opacity =0.1] (0,-1) -- (0,2)--(4,2) -- (4, -1) -- (0,-1) ; 
\draw [very thick] (0,-1) -- (0,2) ;
\draw [very thick] (2,-1) -- (2,2) ;
\draw[thick] (0,1) -- (4,1);
\draw[thick] (0,0.2) -- (4,0.2);
\draw[thick] (0,-0.5) -- (2,-0.5);
\node at (1,1.65) {$\cZ(\cH_3)$} ;
\node at (3,1.65) {$\cZ(\Z_3)$} ;
\node[above] at (0,2) {$\cL_3$}; 
\node[above] at (2,2) {$\cA_1$}; 
\node[below] at (1,1) {$\pi_2$}; 
\node[below] at (3,1) {$e^2$}; 
\node[below] at (1,0.2) {$\pi_2$}; 
\node[below] at (3,0.2) {$e$}; 
\node[below] at (1,-0.5) {$\pi_1$}; 
\draw [black,fill=black] (0,1) ellipse (0.05 and 0.05);
\draw [black,fill=black] (0,0.2) ellipse (0.05 and 0.05);
\draw [black,fill=black] (0,-0.5) ellipse (0.05 and 0.05);
\draw [black,fill=black] (2,1) ellipse (0.05 and 0.05);
\draw [black,fill=black] (2,0.2) ellipse (0.05 and 0.05);
\draw [black,fill=black] (2,-0.5) ellipse (0.05 and 0.05);
\node at (4.4,0.5) {$=$};
\end{scope}
\begin{scope}[shift={(5.7,0)}]
\draw [fill= black, opacity =0.1] (0,-1) -- (0,2)--(2,2) -- (2, -1) -- (0,-1) ; 
\draw [very thick] (0,0) -- (0,2) ;
\draw [very thick] (0,-1) -- (0,2) ;
\node[below] at (1,1) {$e^2$}; 
\node[below] at (1,0.2) {$e$}; 
\node at (1,1.65) {$\cZ(\Z_3)$} ;
\node[above] at (0,2) {$2\cL_e$}; 
\node at (-0.5,1) {$\cE^{1,2}_{e^2}$};
\node at (-0.5,0.2) {$\cE^{1,2}_{e}$};
\node at (-0.5,-0.5) {$\cO$};
\draw[thick] (0,1) -- (2,1);
\draw[thick] (0,0.2) -- (2,0.2);
\draw [black,fill=black] (0,1) ellipse (0.05 and 0.05);
\draw [black,fill=black] (0,0.2) ellipse (0.05 and 0.05);
\draw [black,fill=black] (0,-0.5) ellipse (0.05 and 0.05);
\end{scope}
\end{tikzpicture}
\end{equation}

By collapsing the symmetry boundary of $\cZ(\cH_3)$ with the interface specified by $\cA_1$, we obtain a gapped boundary of $\cZ(\Z_3)$. The Lagrangian algebra defining this gapped boundary can be determined by looking at the lines ending on $\cL_3$ and following what they become in $\cZ(\Z_3)$. From the picture above we see that the boundary of the reduced topological order is 
\begin{equation}
    \fB' = \fB_0^e \oplus \fB_1^e \,,
\end{equation}
i.e.\ the sum of two electric boundary conditions, as specified by the Lagrangian algebra 
\begin{equation}
    \cL_{\fB'} = 2(1\oplus e \oplus e^2) \,.
\end{equation}
In particular, the boundary is reducible. The topological lines living on $\fB'$ form a multi-fusion category which we call $\text{Mat}_2(\Z_3) $. Its simple objects are 
\begin{equation}
    1_{ij} \,, \eta_{ij} \,, \eta^2_{ij} \quad i,j \in \{0,1\} \,.
\end{equation}
In particular, $\{1_{ii},\eta_{ii},\eta^2_{ii} \}$ are the lines generating the $\Z_3$ symmetry on $\fB^e_i$, while the other lines for $i\neq j$ are  boundary changing operators. 

Now let us discuss the boundary operators. Notice we get two possible operators $\cE_{e}^{1,2}$ and $\cE_{e^2}^{1,2}$ because $\pi_2$ ends with multiplicity 2 on the boundary specified by $\cL_3$.
The product of boundary operators is  determined using constraints such as the symmetry action, the product of bulk lines, and associativity. 
The symmetry action is given by 
\begin{equation}\label{eq:bdry_action}
\begin{aligned}
    &\alpha: \qquad \cO \rightarrow \cO \,,\quad \cE_{e^i}^1 \rightarrow \omega \cE_{e^i}^1 \,, \quad \cE_{e^i}^2 \rightarrow \omega^2 \cE_{e^i}^2 \\
    &\rho: \qquad \cO \rightarrow -d^{-1} \cO  \,, \quad \cE_{e^i}^1 \leftrightarrow \cE_{e^i}^2\,.
\end{aligned}
\end{equation}
Using this, we can determine the products of local operators to be 
\begin{equation}
\begin{aligned}
    &\cO^2 = 1 +3\cO \\
    & \cO \cE_e^1 = d \, \cE_e^1 \\
    & \cO \cE_e^2 = -d^{-1} \cE_e^2
\end{aligned}   \qquad \quad
\begin{aligned}
    &\cE_e^1 \cE_e^1 = \cE_{e^2}^2 \\
    &\cE_e^2 \cE_e^2 = \cE_{e^2}^1 \\
    &\cE_e^1 \cE_e^2 = 0 \,.
\end{aligned}
\end{equation}
Notice in particular this implies we can focus on only $\cE_{e}^{1,2}$, as the operators $\cE_{e^2}^{1,2}$ are determined in terms of them. We construct local operators
\begin{equation}
    v_0 = \frac{d^{-1}+\cO}{\sqrt{13}} \quad,\quad v_1 = \frac{d - \cO}{\sqrt{13}} \,.
\end{equation}
We can interpret $v_0$ as the identity local operator along the irreducible boundary $\fB_0^e$, and similarly $v_1$ as the identity along $\fB_1^e$. This follows from the fact that $v_0$ and $v_1$ are orthogonal idempotents, satisfying $v_i v_j = \delta_{ij} v_j$. Moreover, $\cE_e^1$ is the end of the $e$ anyon along $\fB_0^e$, while $\cE_e^2$ is the end of the $e$ anyon along $\fB_1^e$, as we have
\begin{equation}
\begin{aligned}
    &\cE_e^1 v_0 = \cE_e^1 \,, \quad \cE_e^2 v_0 = 0 \,, \quad  \cE_e^1 v_1 = 0 \\
    &\cE_e^2 v_1 = \cE_e^2 \,, \quad  \cE_e^1 \cE_e^2 = 0 \,.
\end{aligned}    
\end{equation}
This implies that indeed $v_0$ ($v_1$) acts as the identity for $\cE_e^1$ ($\cE_e^2$), and that the overlap between the two boundaries is trivial. 

We are now ready to discuss how the Haagerup symmetry acts on the boundary $\fB'$, using \eqref{eq:bdry_action}. We see that $\alpha$ acts within each of the two irreducible boundaries as
\begin{equation}\label{alpha_app}
    \alpha = \eta_{00} \oplus \eta_{11}^2 \,. 
\end{equation}
The non-invertible generator $\rho$ instead maps between the two boundaries, since it sends 
\begin{equation}\label{eq:universe}
    v_0 \rightarrow d^{-1} v_1  \,, \quad v_1 \rightarrow d\, v_0 + 3 v_1 \,,
\end{equation}
while also switching $\cE^1_{e^i}$ with $\cE^2_{e^i}$, from which we deduce
\begin{equation}\label{rho_app}
    \rho = 1_{01} \oplus 1_{10} \oplus 1_{11} \oplus \eta_{11} \oplus \eta^2_{11} ,.
\end{equation}
Notice there is a relative Euler counterterm $e^{-\lambda} = d^{-1}$ between $\fB_0^e$ and $\fB_1^e$.
One can easily check that all the relations defining $\cH_3$ are satisfied, for example the non-commutativity 
\begin{equation}
    \alpha \rho = \eta_{01} \oplus \eta_{10}^2 \oplus 1_{11} \oplus \eta_{11} \oplus \eta_{11}^2 = \rho \alpha^2.  
\end{equation}
This provides mathematically a functor from $\cH_3$
to Mat$_2(\Z_3)$, the category formed by the lines on the reducible gapped boundary $\fB'$.

We are now ready to construct a gapless phase with Haagerup symmetry using the club sandwich.
We first remark that 
\be
\cL_1 \cap \cL_3= \cA_1 \,, 
\ee
so that we expect this gapless phase to be a standard second order phase transitions at the critical point between the 2 vacua and the 6 vacua gapped phases. 

The simplest CFT at the phase transition is the 3-state Potts (3SP) minimal model with central charge $c=4/5$. Here we  focus on the $\Z_3$ symmetry of the 3SP, but it is known this CFT admits a larger fusion category symmetry with 16 simple objects, as analysed in \cite{Chang:2018iay}.
The two possible gapped phases for $\Z_3$ can be obtained by deforming the 3SP by the relevant $\Z_3$ symmetric operator $\epsilon$ of conformal dimension $(2/5, 2/5)$. The CFT flows to one of the two $\Z_3$ phases depending on the sign of the deformation, as discussed e.g. in \cite{Fendley2019}.
In particular, it flows to the disordered ($\Z_3$-trivial) phase for $-\epsilon$ and ordered ($\Z_3$-SSB) phase for $+ \epsilon$. 

We can now feed the 3SP model in as the physical boundary of the club quiche \eqref{CQA1}, which after compactification results in the $\cH_3$-symmetric theory
\be\label{bingo_app}
\begin{tikzpicture}
\node at (1.5,-3) {$\text{\bf 3SP}_0\oplus\text{\bf 3SP}_1$};
\node[right] at (3.2,-3) {$\Z_3, \rho$};
\draw [-stealth](2.7,-3.2) .. controls (3.2,-3.5) and (3.2,-2.6) .. (2.7,-2.9);
\draw [-stealth,rotate=180](-0.3,2.8) .. controls (0.2,2.5) and (0.2,3.4) .. (-0.3,3.1);
\node at (-0.4,-3) {$\Z_3$};
\draw [-stealth](0.8,-3.3) .. controls (1,-3.8) and (2,-3.8) .. (2.2,-3.3);
\node at (1.5,-4) {$\rho$};
\draw [-stealth,rotate=180](-2.2,2.7) .. controls (-2,2.2) and (-1,2.2) .. (-0.8,2.7);
\node at (1.5,-2) {$\rho$};
\end{tikzpicture}
\ee

\noindent{\bf Relevant deformations.}
We now want to show that this CFT admits relevant deformations to the 2 and 6 vacua $\cH_3$ symmetric gapped phases, i.e.\ that it is indeed the phase transition between them.
We start by considering the deformation $\epsilon_0 + \epsilon_1$. Notice that using \eqref{eq:universe} and \eqref{rho_app} we find $\rho(\epsilon_0 + \epsilon_1) = d (\epsilon_0 + \epsilon_1)$, so that the deformation in indeed fully Haagerup symmetric. This comes from the fact that $\epsilon_0$ and $\epsilon_1$ are $\Z_3$ symmetric, so that $\eqref{rho_app}$ becomes $\rho = 1_{01} \oplus 1_{10} \oplus 3 \,1_{11}$, and the use of the Euler terms shown in \eqref{eq:universe}. Adding this relevant deformation with the $-$ sign, both copies of 3SP flow to the trivially symmetric gapped phase. Let us denote the 2 vacua of each trivial phase by $v_0$ and $v_1$ respectively. In each of these, the $\Z_3$ symmetry is completely trivial, which implies 
$\eta_{ii} = 1_{ii}$.
Therefore, using \eqref{alpha_app} and \eqref{rho_app}, the $\cH_3$ symmetry generators descend to 
\begin{equation}
\begin{aligned}
    &\alpha = 1_{00} \oplus 1_{11} \\
    &\rho = 1_{01} \oplus 1_{10} \oplus 3\, 1_{11} \,.
\end{aligned}    
\end{equation}
This precisely reproduces the symmetry action we discussed for the $\cH_3$ symmetric gapped phases with 2 vacua. Now consider the opposite $+$ sign deformation. Both 3SP now flow to the $\Z_3$ SSB phase. Therefore we obtain 3 vacua for each 3SP, and 6 in total. Let us denote by $v_0, v_1, v_2$ the three vacua of the gapped phase to which 3SP$_1$ flows, and by $v_3, v_4, v_5$ the 3 vacua for 3SP$_0$. In this case, the $\Z_3$ symmetry is spontaneously broken and therefore it is represented on the vacua as
\begin{equation}
\begin{aligned}
    &\eta_{11} = 1_{01} \oplus 1_{12} \oplus 1_{20} \\
    &\eta_{00} = 1_{34} \oplus 1_{45} \oplus 1_{53} \,.
\end{aligned}
\end{equation}
Therefore, using \eqref{alpha_app} and \eqref{rho_app}, the symmetry generators can be represented on the six vacua as 
\begin{equation}
\begin{aligned}
    &\alpha = 1_{02} \oplus 1_{12} \oplus 1_{21} \oplus 1_{34} \oplus 1_{45} \oplus 1_{53} \\
    &\rho = 1_{03} \oplus 1_{14} \oplus 1_{25} \oplus 1_{30} \oplus 1_{41} \oplus 1_{52} \oplus 1_{00} \oplus 1_{11} \\
    &\oplus 1_{22} \oplus 1_{01} \oplus 1_{12} \oplus 1_{20} \oplus 1_{02} \oplus 1_{21} \oplus 1_{10} \,.
\end{aligned}    
\end{equation}
This reproduces exactly the symmetry action we discussed for the full $\cH_3$ symmetric gapped phase with 6 vacua. 

Starting with the theory (\ref{bingo_app}), all combinations of relevant deformations are allowed and give rise to the various gapped phases with $n=2, 4, 6$ vacua respectively, labelled by $\Phi_n^{\epsilon_0, \epsilon_1}$. These arise from the relevant deformation of each of the 3SP models to a single or three vacua gapped phase. The transition between $\Phi_2^{--}$ and $\Phi_6^{++}$ is second order, likewise between $\Phi_4^{+-}$ and $\Phi_4^{-+}$. These are all KT transformations of the $\Z_3$-transition between the trivial and SSB phases -- in various combinations. The transitions between $\Phi_4^{+-}$ and $\Phi_6^{++}$ or $\Phi_4^{+-}$ and $\Phi_2^{--}$, etc. as $\Z_3$ transitions
correspond to a gapped phase for one of the two universes, and a second order phase transition in the other. 
From our analysis using the SymTFT, we can only say that the truly $\cH_3$-symmetric second order phase transition is the one between $\Phi_6$ and $\Phi_2$. This does not say anything about the 4 vacua gapped phases. 
However, we see the existence of a deformation starting with the CFT (\ref{bingo_app}) to a 4 vacua gapped phase from turning $\epsilon_0=-\epsilon_1$, which suggestively decomposes ad a $\Z_3$ SSB  and a trivial phase.

\section{Haagerup-symmetric Lattice Models}

To complement the continuum analysis, we now construct UV lattice models with Haagerup symmetry that flow to the gapped and gapless phases we discussed above from the SymTFT perspective. We focus on a class of models known as anyon chains \cite{Feiguin:2006ydp,2009arXiv0902.3275T, Aasen:2016dop, Buican:2017rxc, Aasen:2020jwb, Lootens:2021tet, Lootens:2022avn, Inamura:2021szw}, which can be defined using an input fusion category  that determines naturally the symmetry $\cS$ of the model.

\subsubsection{Anyon Chains for Gapped and Gapless Phases}

Let us recap the main steps of the construction, following \cite{Bhardwaj:2024kvy}. 
The input data entering the definition of the anyon chain model is:
\begin{itemize}
    \item An input fusion category $\cC$, which should in general be distinguished from the symmetry fusion category $\cS$.
    \item A $\cC$-module category $\cM$. The symmetry $\cS$ is determined in terms of $\cC$ and $\cM$ as 
    \be
    \cS=\cC^*_\cM = \text{Fun}_{\cC} (\cM, \cM)\,,
    \ee 
    which is the category formed by $\cC$-module functors from $\cM$ to $\cM$.
\item An object $r$ in $\cC$, which in general is taken to be a non-simple object.
\end{itemize}
The basic constituent for the lattice model is then a block of the following form
\be
\begin{split}
\begin{tikzpicture}
\draw [thick,-stealth](-2,-0.5) -- (-0.25,-0.5);
\draw [thick](-0.5,-0.5) -- (1.5,-0.5);
\begin{scope}[shift={(2,0)}]
\draw [thick,-stealth](-0.5,-0.5) -- (1.25,-0.5);
\draw [thick](1,-0.5) -- (3,-0.5);
\end{scope}
\begin{scope}[shift={(1,0)},rotate=90]
\draw [thick,-stealth](-0.5,-0.5) -- (0.25,-0.5);
\draw [thick](0,-0.5) -- (1,-0.5);
\end{scope}
\node at (-0.25,-1) {$m_i\in\cM$};
\node at (3.25,-1) {$m_{i+1}\in\cM$};
\node[] at (1.5,1.5) {$r\in\cC$};
\draw[fill=black] (1.5,-0.5) ellipse (0.08 and 0.08);
\node[] at (1.5,-1) {$\mu_{i+\half}$};
\end{tikzpicture}
\end{split}
\ee
where $m_i$ and $m_{i+1}$ are simple objects in the module category $\cM$, and $\mu_{i+\half}\in\Hom(m_i,r\ot m_{i+1})$ is a basis vector in the morphism space formed by $r$ ending between $m_i$ and $m_{i+1}$. The full Hilbert space of the model is constructed by concatenating such basic blocks. We typically use periodic boundary conditions and identify $m_1 = m_{L+1}$, where $L$ is the length of the chain. The lattice model naturally possesses a $\cS$ symmetry, as topological lines from $\cS$ can be fused from below using the fact that $\cM$ is also a right module category over $\cS$. 

We can realise all the possible gapped phases for $\cS$ as the ground states of specific Hamiltonians acting on the anyon chain. In particular, let us recall there is a one-to-one correspondence between Lagrangian algebras of the Drinfeld center $\cZ(\cS)$ and Frobenius algebra objects in $\cS$ (they both determine a generalised gauging of all or part of the fusion category symmetry $\cS$). To realise the gapped phase specified by a certain Lagrangian algebra $\cL$, we then pick the corresponding Frobenius algebra object $F$ and consider the operator $H_i^F$ 
\be\label{eq:ham_app}
\begin{split}
\begin{tikzpicture}
\begin{scope}[shift={(6.75,0)}]
\draw [thick,-stealth](-0.5,-0.5) -- (0.25,-0.5);
\draw [thick](0,-0.5) -- (1.5,-0.5);
\node at (0.25,-1) {$m_{i-1}$};
\begin{scope}[shift={(2,0)}]
\draw [thick,-stealth](-0.5,-0.5) -- (0.25,-0.5);
\draw [thick](0,-0.5) -- (1.25,-0.5);
\node at (0.25,-1) {$m_{i}$};
\end{scope}
\begin{scope}[shift={(0.75,0)},rotate=90]
\draw [thick,-stealth,black](-0.5,-0.5) -- (0.25,-0.5);
\draw [thick,black](-0.5,-0.5) -- (0.75,-0.5);
\node[black] at (2.25,-2.5) {$F$};
\node[black] at (1,-1.5) {$F$};
\draw[fill=black] (-0.5,-0.5) ellipse (0.08 and 0.08);
\node[black] at (-1,-0.5) {$\mu_{i-\half}$};
\end{scope}
\begin{scope}[shift={(3.75,0)}]
\draw [thick,-stealth](-0.5,-0.5) -- (0.5,-0.5);
\draw [thick](0.25,-0.5) -- (1,-0.5);
\node at (0.5,-1) {$m_{i+1}$};
\end{scope}
\begin{scope}[shift={(2.75,0)},rotate=90]
\draw [thick,-stealth,black](-0.5,-0.5) -- (0.25,-0.5);
\draw [thick,black](-0.5,-0.5) -- (0.75,-0.5);
\node[black] at (2.25,1.5) {$F$};
\node[black] at (0.25,-0.25) {$F$};
\node[black] at (0.25,1.25) {$F$};
\draw[fill=black] (-0.5,-0.5) node (v1) {} ellipse (0.08 and 0.08);
\node[black] at (-1,-0.5) {$\mu_{i+\half}$};
\end{scope}
\draw [thick,-stealth,black](3.25,1.25) -- (3.25,1.5);
\draw [thick,-stealth,black](1.25,1.25) -- (1.25,1.5);
\draw [thick,-stealth,black](2,0.75) -- (2.25,0.75);
\draw [thick,black](3.25,0.75) -- (3.25,2);
\draw [thick,black](1.25,0.75) -- (1.25,2);
\draw [thick,black](3.25,0.75) -- (1.25,0.75);
\draw[fill=black] (3.25,0.75) ellipse (0.08 and 0.08);
\draw[fill=black] (1.25,0.75) ellipse (0.08 and 0.08);
\node[black] at (3.75,0.75) {$m$};
\node[black] at (0.75,0.75) {$\Delta$};
\end{scope}
\end{tikzpicture}
\end{split}
\ee
acting at site $i$. In the above, $m$ and $\Delta$ are the multiplication and co-multiplication of the Frobenius algebra $F$. The full Hamiltonian is 
\begin{equation}
    H^F = - \sum_j H_j^F \,.
\end{equation}
Notice this intuitively commutes with the symmetry $\cS$, as the symmetry lines are fused from below while the Hamiltonian operator acts from above on the anyon chain. 

Ground states of this Hamiltonian can simply be identified with left modules $\cK$ for the Frobenius algebra $F$. Such a module is a (not necessarily simple) object $m\in\cM$ along with a morphism $\mu\in\Hom(m,F\ot m)$ satisfying a series of properties (see e.g.\ equation (3.7) in \cite{Bhardwaj:2024kvy}) which imply 
\be
\begin{split}
\scalebox{0.75}{\centering
\begin{tikzpicture}
\draw [thick,-stealth](-0.5,-0.5) -- (0.25,-0.5);
\draw [thick](0,-0.5) -- (1.5,-0.5);
\node at (0.25,-1) {$m$};
\begin{scope}[shift={(2,0)}]
\draw [thick,-stealth](-0.5,-0.5) -- (0.25,-0.5);
\draw [thick](0,-0.5) -- (1.25,-0.5);
\node at (0.25,-1) {$m$};
\end{scope}
\begin{scope}[shift={(0.75,0)},rotate=90]
\draw [thick,-stealth,black](-0.5,-0.5) -- (0.25,-0.5);
\draw [thick,black](-0.5,-0.5) -- (0.75,-0.5);
\node[black] at (2.25,-2.5) {$F$};
\node[black] at (1,-1.5) {$F$};
\draw[fill=black] (-0.5,-0.5) ellipse (0.1 and 0.1);
\node[black] at (-1,-0.5) {$\mu$};
\end{scope}
\begin{scope}[shift={(3.75,0)}]
\draw [thick,-stealth](-0.5,-0.5) -- (0.5,-0.5);
\draw [thick](0.25,-0.5) -- (1,-0.5);
\node at (0.5,-1) {$m$};
\end{scope}
\begin{scope}[shift={(2.75,0)},rotate=90]
\draw [thick,-stealth,black](-0.5,-0.5) -- (0.25,-0.5);
\draw [thick,black](-0.5,-0.5) -- (0.75,-0.5);
\node[black] at (2.25,1.5) {$F$};
\node[black] at (0.25,-0.25) {$F$};
\node[black] at (0.25,1.25) {$F$};
\draw[fill=black] (-0.5,-0.5) node (v1) {} ellipse (0.1 and 0.1);
\node[black] at (-1,-0.5) {$\mu$};
\end{scope}
\draw [thick,-stealth,black](3.25,1.25) -- (3.25,1.5);
\draw [thick,-stealth,black](1.25,1.25) -- (1.25,1.5);
\draw [thick,-stealth,black](2,0.75) -- (2.25,0.75);
\draw [thick,black](3.25,0.75) -- (3.25,2);
\draw [thick,black](1.25,0.75) -- (1.25,2);
\draw [thick,black](3.25,0.75) -- (1.25,0.75);
\draw[fill=black] (3.25,0.75) ellipse (0.1 and 0.1);
\draw[fill=black] (1.25,0.75) ellipse (0.1 and 0.1);
\node at (5,0.75) {=};
\begin{scope}[shift={(6,0)}]
\draw [thick,-stealth](-0.5,-0.5) -- (0.25,-0.5);
\draw [thick](0,-0.5) -- (1.5,-0.5);
\node at (0.25,-1) {$m$};
\begin{scope}[shift={(2,0)}]
\draw [thick,-stealth](-0.5,-0.5) -- (0.25,-0.5);
\draw [thick](0,-0.5) -- (1.25,-0.5);
\node at (0.25,-1) {$m$};
\end{scope}
\begin{scope}[shift={(0.75,0)},rotate=90]
\draw [thick,black](-0.5,-0.5) -- (0.75,-0.5);
\node[black] at (2.25,-2.5) {$F$};
\draw[fill=black] (-0.5,-0.5) ellipse (0.1 and 0.1);
\node[black] at (-1,-0.5) {$\mu$};
\end{scope}
\begin{scope}[shift={(3.75,0)}]
\draw [thick,-stealth](-0.5,-0.5) -- (0.5,-0.5);
\draw [thick](0.25,-0.5) -- (1,-0.5);
\node at (0.5,-1) {$m$};
\end{scope}
\begin{scope}[shift={(2.75,0)},rotate=90]
\draw [thick,black](-0.5,-0.5) -- (0.75,-0.5);
\node[black] at (2.25,1.5) {$F$};
\draw[fill=black] (-0.5,-0.5) node (v1) {} ellipse (0.1 and 0.1);
\node[black] at (-1,-0.5) {$\mu$};
\end{scope}
\draw [thick,-stealth,black](3.25,0.5) -- (3.25,0.75);
\draw [thick,-stealth,black](1.25,0.5) -- (1.25,0.75);
\draw [thick,black](3.25,0.75) -- (3.25,2);
\draw [thick,black](1.25,0.75) -- (1.25,2);
\end{scope}
\end{tikzpicture}}
\end{split}
\ee
It follows that a state constructed out of an $F$-module is a $+1$ eigenstate of all projectors $H_j^F$ and hence a ground state.  Notice that in the case $\cM$ is the regular module, $\cK$ becomes a standard left module over the algebra $F$ \cite{Inamura:2021szw}
\begin{equation}
    \cK \in \Mod_\cC(F) \,.
\end{equation} 

We now apply this logic to construct the Haagerup symmetric gapped phases from the anyon chain:
the choice of input data is 
\be
\cC= \cM = \cS= \cH_3 \,,
\ee
i.e. the module category $\cM$ is the regular module for $\cC$ and this gives rise to the symmetry $\cS = \cC^*_\cM = \cH_3$. 
The Hilbert space of the model on a lattice of length $L$ with periodic boundary conditions is spanned by states corresponding to the fusion trees 
\be\label{eq:state}
\begin{split}
\begin{tikzpicture}
\draw [thick,-stealth](0,-0.5) -- (0.75,-0.5);
\draw [thick](0.5,-0.5) -- (1.5,-0.5);
\node at (0.75,-1) {$m_1$};
\begin{scope}[shift={(2,0)}]
\draw [thick,-stealth](-0.5,-0.5) -- (0.25,-0.5);
\draw [thick](0,-0.5) -- (1,-0.5);
\node at (0.25,-1) {$m_{2}$};
\end{scope}
\begin{scope}[shift={(1,0)},rotate=90]
\draw [thick,-stealth](-0.5,-0.5) -- (0.25,-0.5);
\draw [thick](0,-0.5) -- (0.75,-0.5);
\node[] at (1,-0.5) {$r$};
\draw[fill=black] (-0.5,-0.5) ellipse (0.08 and 0.08);
\node[] at (-1,-0.5) {$\mu_{\frac32}$};
\end{scope}
\begin{scope}[shift={(2.5,0)},rotate=90]
\draw [thick,-stealth](-0.5,-0.5) -- (0.25,-0.5);
\draw [thick](0,-0.5) -- (0.75,-0.5);
\node[] at (1,-0.5) {$r$};
\draw[fill=black] (-0.5,-0.5) node (v1) {} ellipse (0.08 and 0.08);
\node[] at (-1,-0.5) {$\mu_{\frac52}$};
\end{scope}
\node (v2) at (4,-0.5) {$\cdots$};
\draw [thick] (v1) edge (v2);
\begin{scope}[shift={(5.5,0)}]
\draw [thick,-stealth](-0.5,-0.5) -- (0.5,-0.5);
\draw [thick](0.25,-0.5) -- (1.25,-0.5);
\node at (0.5,-1) {$m_{L}$};
\end{scope}
\begin{scope}[shift={(4.5,0)},rotate=90]
\draw [thick,-stealth,black](-0.5,-0.5) -- (0.25,-0.5);
\draw [thick](0,-0.5) -- (0.75,-0.5);
\node[] at (1,-0.5) {$r$};
\draw[fill=black] (-0.5,-0.5) node (v1) {} ellipse (0.08 and 0.08);
\node[] at (-1,-0.5) {$\mu_{L-\half}$};
\end{scope}
\draw [thick] (v2) edge (v1);
\begin{scope}[shift={(7.25,0)}]
\draw [thick,-stealth](-0.5,-0.5) -- (0.25,-0.5);
\draw [thick](0,-0.5) -- (0.75,-0.5);
\node at (0.25,-1) {$m_{1}$};
\end{scope}
\begin{scope}[shift={(6.25,0)},rotate=90]
\draw [thick,-stealth](-0.5,-0.5) -- (0.25,-0.5);
\draw [thick](0,-0.5) -- (0.75,-0.5);
\node[] at (1,-0.5) {$r$};
\draw[fill=black] (-0.5,-0.5) node (v1) {} ellipse (0.08 and 0.08);
\node[] at (-1,-0.5) {$\mu_{\half}$};
\end{scope}
\end{tikzpicture}
\end{split}
\ee
where
\be
r = \bigoplus_{l \in \cH_3} l \,.
\ee
Note this element is usually called $\rho$, which is not a suitable notation in the present context.

We can construct three commuting projector Hamiltonians whose ground states realise each of the three Haagerup symmetric gapped phases. These are labelled by Frobenius algebras in the input category $\cC = \cH_3$, which are
\be
F_1 = 1 \,,\qquad F_2 = 1 \oplus  \alpha \oplus \alpha^2 \,,\qquad F_3 = 1 \oplus \rho \oplus \alpha \rho \,.
\ee
In particular, we find the following:
\begin{itemize}
    \item $F_1$ $\Leftrightarrow$ $\cH_3$ SSB phase (6 vacua)
    \item $F_2$ $\Leftrightarrow$ $\cH_3/ \Z_3$ SSB phase (2 vacua)
    \item $F_3$ $\Leftrightarrow$ $\Z_3$ SSB $\oplus$ trivial phase (4 vacua) \,.
\end{itemize}

\subsection{Phase Transitions for $\cH_3$}
To realise the phase transition between the 6 vacua case and the 2 vacua case, we use the input choices 
\begin{equation}
    \cC = \cM = \cS = \cH_3 \,,
\end{equation}
which remain the same as for the gapped phases, but we restrict $r$ to be 
\begin{equation}
    r = 1 \oplus \alpha \oplus \alpha^2 \,.
\end{equation}
Restricting to this $r$ leads to a direct sum decomposition of the original Hilbert space into state spaces $V_0$ and $V_1$ spanned by 
\be
\begin{split}
\begin{tikzpicture}
\draw [thick,-stealth](0,-0.5) -- (0.75,-0.5);
\draw [thick](0.5,-0.5) -- (1.5,-0.5);
\node at (0.75,-1) {\footnotesize $1,\alpha,\alpha^2$};
\begin{scope}[shift={(2,0)}]
\draw [thick,-stealth](-0.5,-0.5) -- (0.25,-0.5);
\draw [thick](0,-0.5) -- (1,-0.5);
\node at (0.25,-1) {\footnotesize $1,\alpha,\alpha^2$};
\end{scope}
\begin{scope}[shift={(1,0)},rotate=90]
\draw [thick,-stealth](-0.5,-0.5) -- (0.25,-0.5);
\draw [thick](0,-0.5) -- (0.75,-0.5);
\node[] at (1,-0.5) {$r$};
\draw[fill=black] (-0.5,-0.5) ellipse (0.08 and 0.08);
\end{scope}
\begin{scope}[shift={(2.5,0)},rotate=90]
\draw [thick,-stealth](-0.5,-0.5) -- (0.25,-0.5);
\draw [thick](0,-0.5) -- (0.75,-0.5);
\node[] at (1,-0.5) {$r$};
\draw[fill=black] (-0.5,-0.5) node (v1) {} ellipse (0.08 and 0.08);
\end{scope}
\node (v2) at (4,-0.5) {$\cdots$};
\draw [thick] (v1) edge (v2);
\begin{scope}[shift={(5.5,0)}]
\draw [thick,-stealth](-0.5,-0.5) -- (0.5,-0.5);
\draw [thick](0.25,-0.5) -- (1.25,-0.5);
\node at (0.5,-1) {\footnotesize $1,\alpha,\alpha^2$};
\end{scope}
\begin{scope}[shift={(4.5,0)},rotate=90]
\draw [thick,-stealth,black](-0.5,-0.5) -- (0.25,-0.5);
\draw [thick](0,-0.5) -- (0.75,-0.5);
\node[] at (1,-0.5) {$r$};
\draw[fill=black] (-0.5,-0.5) node (v1) {} ellipse (0.08 and 0.08);
\end{scope}
\draw [thick] (v2) edge (v1);
\begin{scope}[shift={(7.25,0)}]
\draw [thick,-stealth](-0.5,-0.5) -- (0.25,-0.5);
\draw [thick](0,-0.5) -- (0.75,-0.5);
\node at (0.25,-1) {\footnotesize $1,\alpha,\alpha^2$};
\end{scope}
\begin{scope}[shift={(6.25,0)},rotate=90]
\draw [thick,-stealth](-0.5,-0.5) -- (0.25,-0.5);
\draw [thick](0,-0.5) -- (0.75,-0.5);
\node[] at (1,-0.5) {$r$};
\draw[fill=black] (-0.5,-0.5) node (v1) {} ellipse (0.08 and 0.08);
\end{scope}
\end{tikzpicture}
\end{split}
\ee
and 
\be\label{eq:state}
\begin{split}
\begin{tikzpicture}
\draw [thick,-stealth](0,-0.5) -- (0.75,-0.5);
\draw [thick](0.5,-0.5) -- (1.5,-0.5);
\node at (0.75,-1) {\footnotesize $\rho,\alpha \rho,\alpha^2 \rho$};
\begin{scope}[shift={(2,0)}]
\draw [thick,-stealth](-0.5,-0.5) -- (0.25,-0.5);
\draw [thick](0,-0.5) -- (1,-0.5);
\node at (0.25,-1) {\footnotesize $\rho,\alpha \rho,\alpha^2 \rho$};
\end{scope}
\begin{scope}[shift={(1,0)},rotate=90]
\draw [thick,-stealth](-0.5,-0.5) -- (0.25,-0.5);
\draw [thick](0,-0.5) -- (0.75,-0.5);
\node[] at (1,-0.5) {$r$};
\draw[fill=black] (-0.5,-0.5) ellipse (0.08 and 0.08);
\end{scope}
\begin{scope}[shift={(2.5,0)},rotate=90]
\draw [thick,-stealth](-0.5,-0.5) -- (0.25,-0.5);
\draw [thick](0,-0.5) -- (0.75,-0.5);
\node[] at (1,-0.5) {$r$};
\draw[fill=black] (-0.5,-0.5) node (v1) {} ellipse (0.08 and 0.08);
\end{scope}
\node (v2) at (4,-0.5) {$\cdots$};
\draw [thick] (v1) edge (v2);
\begin{scope}[shift={(5.5,0)}]
\draw [thick,-stealth](-0.5,-0.5) -- (0.5,-0.5);
\draw [thick](0.25,-0.5) -- (1.25,-0.5);
\node at (0.5,-1) {\footnotesize $\rho,\alpha \rho,\alpha^2 \rho$};
\end{scope}
\begin{scope}[shift={(4.5,0)},rotate=90]
\draw [thick,-stealth,black](-0.5,-0.5) -- (0.25,-0.5);
\draw [thick](0,-0.5) -- (0.75,-0.5);
\node[] at (1,-0.5) {$r$};
\draw[fill=black] (-0.5,-0.5) node (v1) {} ellipse (0.08 and 0.08);
\end{scope}
\draw [thick] (v2) edge (v1);
\begin{scope}[shift={(7.25,0)}]
\draw [thick,-stealth](-0.5,-0.5) -- (0.25,-0.5);
\draw [thick](0,-0.5) -- (0.75,-0.5);
\node at (0.25,-1) {\footnotesize $\rho,\alpha \rho,\alpha^2 \rho$};
\end{scope}
\begin{scope}[shift={(6.25,0)},rotate=90]
\draw [thick,-stealth](-0.5,-0.5) -- (0.25,-0.5);
\draw [thick](0,-0.5) -- (0.75,-0.5);
\node[] at (1,-0.5) {$r$};
\draw[fill=black] (-0.5,-0.5) node (v1) {} ellipse (0.08 and 0.08);
\end{scope}
\end{tikzpicture}
\end{split}
\ee
We notice these correspond to two anyon chains defined using the input data $\cC' = \cM' = \Z_3$, which have $\cS' = \Z_3$ symmetry.
Correspondingly, the original module category $\cM = \cH_3$ decomposes as a $\Z_3$ module category as 
\begin{equation}
    \cM \cong \Z_3 \oplus \Z_3 \,.
\end{equation}
Both $V_0$ and $V_1$ are therefore tensor product spaces of local qutrits $|q_i \rangle_{0,1}$ assigned to integer sites, where $q_i = 0,1,2$ depending on whether $m_i = 1,\alpha,\alpha^2$  for $V_0$ and $m_i = \rho,\alpha \rho,\alpha^2 \rho$ for $V_1$ respectively. We denote a basis state as $|\vec{q}\rangle$.

We now consider the Hamiltonian given by 
\begin{equation}\label{eq:SP1_lattice}
    \cH = - \sum_j \left[ \begin{tikzpicture}[scale=0.5, baseline=(current bounding box.center)]
\scriptsize{
\draw[thick, ->-,] (-1,0) -- (-1,1);
\draw[thick, ->-,] (-1,1) -- (-1,2);
\draw[thick, ->-,] (-1,1) -- (2,1);
\draw[thick, ->-,] (2,0) -- (2,1);
\draw[thick, ->-,] (2,1) -- (2,2);
\draw[thick,fill=black] (-1,1) ellipse (0.1 and 0.1);
\draw[thick,fill=black] (2,1) ellipse (0.1 and 0.1);
\node[] at (0.4, 1.5) {\scriptsize $1$};
\node[] at (-0.4, 0.2) {\scriptsize $1$};
\node[] at (1.4, 0.2) {\scriptsize $1$};
}
\end{tikzpicture} 
+ \frac{\lambda}{3} \sum_{\substack{h,h_L,h_R \in \\  \{ 1,\alpha,\alpha^2 \}}}
\begin{tikzpicture}[scale=0.5, baseline=(current bounding box.center)]
\scriptsize{
\draw[thick, ->-,] (-1,0) -- (-1,1);
\draw[thick, ->-,] (-1,1) -- (-1,2);
\draw[thick, ->-,] (-1,1) -- (2,1);
\draw[thick, ->-,] (2,0) -- (2,1);
\draw[thick, ->-,] (2,1) -- (2,2);
\draw[thick,fill=black] (-1,1) ellipse (0.1 and 0.1);
\draw[thick,fill=black] (2,1) ellipse (0.1 and 0.1);
\node[] at (0.4, 1.5) {\scriptsize $h$};
\node[] at (-0.4, 0.2) {\scriptsize $h_L$};
\node[] at (1.4, 0.2) {\scriptsize $h_R$};
}
\end{tikzpicture}
\right]_j \,,
\end{equation}
where the two terms inside the braket correspond to \eqref{eq:ham_app} with $F=1$ and $F=1\oplus \alpha \oplus \alpha^2$ respectively. Written in a more familiar language, the above Hamiltonian reads
\begin{equation}
\begin{aligned}
    \cH = - &\sum_j  \left( \frac{1+ Z_{j-1}Z_j^\dagger + Z_{j-1}^\dagger Z_{j}}{3}\right) \times  \\
    &\times \left( \frac{1+ Z_{j}Z_{j+1}^\dagger + Z_{j}^\dagger Z_{j+1}}{3}\right) + \lambda \left(\frac{1+X_j + X_j^2}{3} \right) \,,
\end{aligned}    
\end{equation}
where $Z_j$ and $X_j$ are the standard $\Z_3$ clock and shift operators
\begin{equation}
\begin{aligned}
    Z_j | q_j \rangle &= \omega^j  | q_j \rangle \,, \quad \omega = e^{2\pi i / 3 } \\
    X_j | q_j \rangle &=  | q_j +1 \text{ mod } 3\rangle \,.
\end{aligned}    
\end{equation}
Since all the building blocks of this Hamiltonian decompose into mutually commuting projectors, we can instead study the simpler Hamiltonian
\begin{equation}
    \cH_{3\SP} = -\frac{1}{3} \sum_j \left( 1+ Z_{j}Z_{j+1}^\dagger + Z_{j}^\dagger Z_{j+1}\right) + \lambda \left( 1+X_j + X_j^2 \right)
\end{equation}
This is the usual quantum three-state Potts (3SP) model Hamiltonian (up to a shift).
$\cH_{3\SP}$ realizes a $\Z_3$ symmetric trivial phase, a $\Z_3$ SSB phase and the 3SP CFT at $\lambda =1$ \cite{Mong:2014ova}, giving a transition between the two phases. On $V_0 \oplus V_1$, it acts block-diagonally, so that we obtain two decoupled sectors, with $\lambda = 1$ realising the CFT
\begin{equation}
    3\SP_0 \oplus 3\SP_1 \,.
\end{equation}
Let us now compute the $\cH_3$ symmetry action, recalling that this is obtained by fusing a symmetry line from below on the chain. The invertible generator $\alpha$ clearly acts within each space as a $\Z_3$ shift symmetry, i.e.\ $\alpha = \eta_{00} \oplus \eta_{11}^2$, where $\eta = \prod_j X_j$, and the subscript indicates whether it acts on $V_0$ or $V_1$. The non-invertible generator $\rho$ has a more interesting action, as it maps between $V_0$ and $V_1$, as well as acting within $V_1$, as $\rho = 1_{01} \oplus 1_{10} \oplus 1_{11} \oplus \eta_{11}\oplus \eta^2_{11}$. Here $1_{ij}$, $i,j=0,1$, sends a state $|\vec{q}\rangle_i$ in $V_i$ to the exact same corresponding state $|\vec{q}\rangle_j$ in $V_j$.

There are two possible Frobenius algebras we can consider given our choice of $r$, namely $F'_1 = 1$ and $F'_2 = 1\oplus \alpha \oplus \alpha^2$. The Hamiltonian corresponding to $F'_1$ acting on $V_0 \oplus V_1$ has 6 ground states 
\begin{equation}
\begin{aligned}
    |GS, \alpha^i \rangle &= |\alpha^i, \cdots, \alpha^i \rangle \,, \qquad i=0,1,2 \\
    |GS, \alpha^i \rho \rangle &= |\alpha^i \rho, \cdots, \alpha^i \rho \rangle \,, \qquad i=0,1,2  \,.
\end{aligned}    
\end{equation}
$\alpha$ clearly permutes the first set of vacua among themselves
\begin{equation}
\begin{aligned}
    \alpha: \quad &|GS,\alpha^i \rangle \rightarrow |GS,\alpha^{i+1} \rangle \\ 
    &|GS,\alpha^i \rho \rangle \rightarrow |GS,\alpha^{i-1} \rho \rangle \,,
\end{aligned}    
\end{equation}
while $\rho$ acts between the two sets of ground states
\begin{equation}
\begin{aligned}
    \rho: \quad &|GS,\alpha^i \rangle \rightarrow |GS,\alpha^i \rho \rangle \\ 
    &|GS,\alpha^i \rho \rangle \rightarrow |GS,\alpha^i  \rangle + \sum_{j=0}^2 |GS,\alpha^j \rho \rangle \,.
\end{aligned}    
\end{equation}
This fully reproduces the continuum results \eqref{eq:alpha_6vacua} and \eqref{eq:rho_6vacua}, therefore giving the $\cH_3$ SSB phase.
The Hamiltonian corresponding to $F'_2$ acting on $V_0 \oplus V_1$ has 2 ground states 
\begin{equation}
\begin{aligned}
    |GS, 0 \rangle &= \frac{1}{3^{L/2}}\sum_{\vec{g}} | \vec{g} \rangle \,, \qquad g_i= \{ 1,\alpha,\alpha^2 \} \\
    |GS, 1  \rangle &= \frac{1}{3^{L/2}}\sum_{\vec{g}} | \vec{g} \rangle \,, \qquad g_i= \{ \rho,\alpha \rho ,\alpha^2 \rho \}
\end{aligned}    
\end{equation}
These are clearly left invariant by $\alpha$, and are permuted by $\rho$ as 
\begin{equation}
    \rho: \quad |GS,0 \rangle \rightarrow |GS,1 \rangle \;,\;  |GS, 1 \rangle \rightarrow |GS,0 \rangle + 3 |GS, 1 \rangle \,,
\end{equation}
reproducing \eqref{eq:2vacua_alpha}, \eqref{eq:2vacua_rho} and thus giving the $\Z_3$ unbroken phase. 
The model \eqref{eq:SP1_lattice} then realises the Haagerup 6 vacua phase for $\lambda < 1$ and the Haagerup 2 vacua phase for $\lambda > 1$: 
\begin{equation}
\begin{tikzpicture}[scale=0.6, baseline=(current bounding box.center)]
\begin{scope}
\draw[thick,->] (0,0) -- (12,0);
\draw[thick,fill=black] (6,0) ellipse (0.15 and 0.15);
\node at (12.5,0) {\footnotesize $\lambda$};
\node at (6,0.5) {\footnotesize $3\SP_0 \oplus 3\SP_1$};
\node at (10.5,0.5) {\footnotesize Two vacua};
\node at (10.5,-0.5) {\footnotesize $(\rm Triv\oplus Triv)$};
\node at (1.5,0.5) {\footnotesize Six vacua};
\node at (1.5,-0.5) {\footnotesize $(\rm SSB_{
\Z_3}\oplus SSB_{\Z_3})$};
\node at (6,-0.5) {\footnotesize $\lambda=1$};
\end{scope}
\end{tikzpicture}
\end{equation}
This confirms the continuum SymTFT analysis that the phase transition between the two gapped phases with Haagerup symmetry $\Phi_6^{++}$ and $\Phi_2^{--}$ is indeed the CFT given by (\ref{bingo_app}).

\end{document}